\documentclass[12pt]{article}

\usepackage{amssymb}

\usepackage[dvips]{graphicx}

\begin{document}

\title
{Higher derivative regularization and quantum corrections in $N=1$
supersymmetric theories.}

\author{A.B.Pimenov, E.S.Shevtsova,\\
{\small{\em Moscow State University, physical faculty,
department of theoretical physics,}}\\
{\small{\em $117234$, Moscow, Russia}}
\\
\\
A.A.Soloshenko,\\
{\small{\em Joint Institute for Nuclear Research}},\\
{\small{\em Bogoliubov Laboratory of Theoretical Physics}},\\
{\small{\em $141980$, Dubna, Moscow region, Russia}}
\\
\\
K.V.Stepanyantz\thanks{E-mail:$stepan@theor.phys.msu.su$}\\
{\small{\em Moscow State University, physical faculty,
department of theoretical physics,}}\\
{\small{\em $117234$, Moscow, Russia}}}

\maketitle

\begin{abstract}
We review some results of applying the higher covariant derivative
regularization to the investigation of quantum corrections
structure in $N=1$ supersymmetric theories. In particular, we
demonstrate that all integrals, defining the Gell-Mann--Low
function in supersymmetric theories, are integrals of total
derivatives. As a consequence, there is an identity for Green
functions, which does not follow from any known symmetry of the
theory, in $N=1$ supersymmetric theories. We also discuss how to
derive the exact $\beta$-function by methods of the perturbation
theory.
\end{abstract}

\tableofcontents

%%%%%%%%%%%%%%%%%%%%%%%%%%%%%%%%%%%%%%%%%%%%%%%%%%%%%%%%%%%%%%%%%%%%%%%%%%

\section{Introduction.}
\hspace{\parindent}

Supersymmetry is certainly one of the most prominent achievements
of the high energy physics. Soon after its discovery it was found
that in supersymmetric theories the ultraviolet behavior was
essentially improved due to some non-renormalization theorems. For
example, there are no divergences in the $N=4$ supersymmetric
Yang--Mills theory, and in theories with $N=2$ supersymmetry
divergences are present only in the one-loop approximation. That
is why such models are very attractive from the theoretical point
of view. But, possibly, the most interesting fact is an indirect
experimental proof of supersymmetry existence in the Standard
model. It was obtained by precise measuring three coupling
constants and investigating their evolution by the renormgroup
equations. The result is that only in the supersymmetric version
of the Standard model the coupling constants coincide at some
energy scale, which should follow from Grand Unification theories.
So, there is no doubt that investigating supersymmetric theories
and, in particular, their quantum properties, is very interesting.
Dynamics of supersymmetric theories is highly nontrivial. It is
worth mentioning summation of instanton corrections in the $N=2$
supersymmetric Yang--Mills theory \cite{SW} (see also
\cite{M1,M2,M3,M4}) or finding a relation between the $N=4$
supersymmetric Yang--Mills theory and a string theory,
compactified on the manifold $AdS_5\times S_5$, known as the
AdS/CFT-correspondence \cite{Maldacena}.

In this paper we will consider theories with unextended
supersymmetry. Such models are especially interesting, because the
physics seems to be $N=1$ supersymmetric at energies of the order
$10^3$ GeV. Dynamics of $N=1$ supersymmetric theories also has
some interesting features. For example, as a result of
investigating instanton contributions structure, in Ref.
\cite{NSVZ_Instanton} form of the $\beta$-function was suggested
exactly to all orders of the perturbation theory. This
$\beta$-function, called the exact
Novikov--Shifman--Vainshtein--Zakharov (NSVZ) $\beta$-function is

\begin{equation}\label{NSVZ_Beta}
\beta(\alpha) = - \frac{\alpha^2\Big[3 C_2 -
C(R)\Big(1-\gamma(\alpha)\Big)\Big]}{2\pi(1- C_2\alpha/2\pi)},
\end{equation}

\noindent where $\gamma(\alpha)$ is the anomalous dimension of the
matter superfield in a representation $R$, and $C(R)$ is defined
by

\begin{equation}\label{C(R)}
\mbox{tr}\,(T^a T^b) = C(R)\,\delta^{ab}.
\end{equation}

\noindent Such a $\beta$-function has not yet been derived by
methods of the perturbation theory. Its numerous verifications by
explicit calculations up to the four-loop approximation were made
in Refs. \cite{ThreeLoop1,ThreeLoop2,ThreeLoop3}. The authors
calculated the $\beta$-function, defined by divergence in the
$\overline{MS}$-scheme. The main result is the following: if a
subtraction scheme is tuned by a special way, then it is possible
to obtain the exact NSVZ $\beta$-function. Nevertheless, there is
an open question, in which scheme one can obtain this
$\beta$-function. The answer to this question is, in particular,
given in this paper. Moreover, we show that there are new
identities, relating Green functions in some $N=1$ supersymmetric
theories, which do not follow from known symmetries of the theory.
Possibly, this assumes existence of some new invariances. Note
that possibility of their existence was discussed, for example, in
finite $N=1$ supersymmetric theories \cite{Ermushev}. Thus, the
dynamics of $N=1$ supersymmetric theories is highly nontrivial and
deserves further investigation. An important constituent of this
investigation is a regularization \cite{JackJones}. The matter is
that the dimensional regularization \cite{tHooft} breaks the
supersymmetry and is not convenient for studying supersymmetric
theories. Most calculations were made with the dimensional
reduction \cite{Siegel}. It is a modification of the dimensional
regularization, which does not break the supersymmetry explicitly.
But this regularization appears to be inconsistent \cite{Siegel2},
and its using can lead to some artifacts. A consistent
regularization, which does not break the supersymmetry, is the
higher covariant derivative regularization \cite{Slavnov}. Despite
of these attractive features, its using is technically
complicated. That is why before recently this regularization was
applied only once, for the one-loop calculation in the
(non-supersymmetric) Yang--Mills theory \cite{Ruiz}. Taking into
account comments, made in subsequent papers
\cite{Asorey,Bakeyev,PhysLett}, the result of the calculation
coincided with the the standard expression for the one-loop
$\beta$-function (although in original paper \cite{Ruiz} the
authors affirm that it is not so). (For other applications of
higher derivatives see, for example, \cite{Morozov} and the
references therein.)

In this paper we will try to demonstrate that using the higher
covariant derivative regularization allows revealing some
interesting features of the quantum correction structure in $N=1$
supersymmetric theories, which was not known earlier.

This paper is organized as follows.

In Sec. \ref{Section_SUSY} we recall basic information about the
$N=1$ supersymmetric Yang--Mills theory, the background field
method, and the higher covariant derivative regularization in
supersymmetric theories. Then, in Sec. \ref{Section_Calculations}
we present results of explicit calculations, made in
supersymmetric theories with the higher covariant derivative
regularization, and analyze their features. After that, in Sec.
\ref{Section_SD} we try to explain them using the Schwinger--Dyson
equations and Slavnov--Taylor identities. It turns out that in
order to explain results of explicit calculation and to obtain the
exact NSVZ $\beta$-function it is necessary to propose the
existence of a new identity for Green functions. Different forms
of this identity are discussed in Sec. \ref{Section_New_Identity}.
In order to verify this identity, some calculations are made in
the three- and four-loop approximations. They are described in
Sec. \ref{Section_Verification}. Structure of quantum corrections
in supersymmetric theories, obtained with the higher derivative
regularization, is discussed in the Conclusion.

%%%%%%%%%%%%%%%%%%%%%%%%%%%%%%%%%%%%%%%%%%%%%%%%%%%%%%%%%%%%%%%%%%%

\section{$N=1$ supersymmetric Yang--Mills theory, background field method,
and higher derivative regularization} \label{Section_SUSY}
\hspace{\parindent}

In this paper we consider the $N=1$ supersymmetric Yang-Mills
theory with matter fields, which is described in the superspace by
the action

\begin{eqnarray}\label{SYM_Action}
&& S = \frac{1}{2 e^2} \mbox{Re}\,\mbox{tr}\int
d^4x\,d^2\theta\,W_a C^{ab} W_b + \frac{1}{4}\int d^4x\,
d^4\theta\, \Big(\phi^+ e^{2V}\phi
+\tilde\phi^+ e^{-2V^{t}}\tilde\phi\Big) +\nonumber\\
&& + \frac{1}{2}m\int d^4x\, d^2\theta\,\tilde\phi^t\,\phi +
\frac{1}{2}m\int d^4x\, d^2\bar\theta\,\tilde\phi^+\phi^*.
\end{eqnarray}

\noindent Here $\phi$ and $\tilde\phi$ are chiral matter
superfields and $V$ is a real scalar superfield, which contains
the gauge field $A_\mu$ as a component. The superfield $W_a$ is a
supersymmetric analogue of the gauge field stress tensor. It is
defined by

\begin{equation}
W_a = \frac{1}{8} \bar D^2 \Big[e^{-2V} D_a e^{2V}\Big],
\end{equation}

\noindent where $D_a$ and $\bar D_a$ are the right and left
supersymmetric covariant derivatives respectively. In our notation
the gauge superfield $V$ is decomposed with respect to the
generators of a gauge group $T^a$ as $V = e\, V^a T^a$, where $e$
is a coupling constant. The generators of the fundamental
representation we will denote by $t^a$. They are normalized by the
condition

\begin{equation}
\mbox{tr}(t^a t^b) = \frac{1}{2} \delta^{ab}.
\end{equation}

Action (\ref{SYM_Action}) is invariant under the gauge
transformations

\begin{equation}
\phi \to e^{i\Lambda}\phi;\qquad \tilde\phi \to
e^{-i\Lambda^t}\tilde\phi;\qquad e^{2V} \to e^{i\Lambda^+} e^{2V}
e^{-i\Lambda},
\end{equation}

\noindent where $\Lambda$ is an arbitrary chiral superfield. Such
a transformation law means that if the field $\phi$ is in the
representation $R$ of the gauge group $G$, then the field
$\tilde\phi$ is in the representation $\bar R$, conjugated to $R$.

Note that theory (\ref{SYM_Action}) is not the most general
renormalizable supersymmetric model. In principle, it is possible
to consider a case, in which matter superfields are in an
arbitrary representation of a gauge group (instead of $R+\bar R$),
and add terms, cubic in the matter superfields. However,
calculations in model (\ref{SYM_Action}) are simpler. That is why
here we investigate only such a theory.

For quantization of this model it is convenient to use the
background field method. The matter is that the background field
method allows calculating the effective action without manifest
breaking of the gauge invariance. In the supersymmetric case it
can formulated as follows \cite{West, Superspace}: Let us make a
substitution

\begin{equation}\label{Substitution}
e^{2V} \to e^{2V'} \equiv e^{\mbox{\boldmath${\scriptstyle
\Omega}$}^+} e^{2V} e^{\mbox{\boldmath${\scriptstyle \Omega}$}}
\end{equation}

\noindent in action (\ref{SYM_Action}), where
$\mbox{\boldmath${\Omega}$}$ is a background scalar superfield. An
expression for $V'$ is a complicated nonlinear function of $V$,
$\mbox{\boldmath$\Omega$}$, and $\mbox{\boldmath$\Omega^+$}$. We
do not interested in explicit form of this function:

\begin{equation}
V' = V'[V,\mbox{\boldmath$\Omega$}].
\end{equation}

\noindent (For brevity of notation we do not explicitly write the
dependence on $\mbox{\boldmath$\Omega$}^+$ here and below.) The
obtained theory will be invariant under the background gauge
transformations

\begin{eqnarray}\label{Background_Transformations}
&& \phi \to e^{i\Lambda}\phi;\qquad \tilde\phi \to
e^{-i\Lambda^t}\tilde\phi;\qquad V \to e^{iK} V e^{-iK};
\nonumber\\
&& e^{\mbox{\boldmath${\scriptstyle \Omega}$}} \to e^{iK}
e^{\mbox{\boldmath${\scriptstyle \Omega}$}} e^{-i\Lambda};\qquad
e^{\mbox{\boldmath${\scriptstyle \Omega}$}^+} \to e^{i\Lambda^+}
e^{\mbox{\boldmath${\scriptstyle \Omega}$}^+} e^{-iK},
\end{eqnarray}

\noindent where $K$ is an arbitrary real superfield. However,
there is one more invariance. In order to construct it, we define
first the background chiral covariant derivatives

\begin{equation}
\mbox{\boldmath$D$}_a \equiv e^{-\mbox{\boldmath${\scriptstyle
\Omega}$}^+} D_a e^{\mbox{\boldmath${\scriptstyle
\Omega}$}^+};\qquad \bar{\mbox{\boldmath$D$}}_a \equiv
e^{\mbox{\boldmath${\scriptstyle \Omega}$}} \bar D_a
e^{-\mbox{\boldmath${\scriptstyle \Omega}$}}.
\end{equation}

\noindent Acting on some field $X$, which is transformed as $X \to
e^{iK} X$, these covariant derivatives are transformed in the same
way. It is also possible to define a covariant derivative with the
Lorentz index

\begin{equation}
\mbox{\boldmath$D$}_\mu \equiv - \frac{i}{4} (C\gamma^\mu)^{ab}
\Big\{\mbox{\boldmath$D$}_a,\bar{\mbox{\boldmath$D$}}_b\Big\},
\end{equation}

\noindent which will have the same property. It is easy to see
that the action is also invariant under the quantum
transformations

\begin{eqnarray}\label{Quantum_Transformations}
&& e^{\mbox{\boldmath${\scriptstyle \Omega}$}} \phi \to
e^{i\lambda}e^{\mbox{\boldmath${\scriptstyle \Omega}$}}
\phi;\qquad e^{-\mbox{\boldmath${\scriptstyle
\Omega}$}^t}\tilde\phi\to
e^{-i\lambda^t}e^{-\mbox{\boldmath${\scriptstyle \Omega}$}^t}
\tilde\phi;\qquad e^{2V} \to e^{i\lambda^+} e^{2V}
e^{-i\lambda};\nonumber\\
&& \qquad\qquad\qquad\qquad \mbox{\boldmath${\Omega}$} \to
\mbox{\boldmath${\Omega}$};\qquad \mbox{\boldmath${\Omega}$}^+ \to
\mbox{\boldmath${\Omega}$}^+
\end{eqnarray}

\noindent where $\lambda$ is an arbitrary background chiral field,
which satisfies the condition $\bar{\mbox{\boldmath$D$}}\lambda =
0$. Such a superfield can be presented as $\lambda =
e^{\mbox{\boldmath${\scriptstyle \Omega}$}} \Lambda
e^{-\mbox{\boldmath${\scriptstyle \Omega}$}} $, where $\Lambda$ is
a usual chiral superfield.

It is easy to see that after substitution (\ref{Substitution})
action (\ref{SYM_Action}) will be

\begin{eqnarray}\label{Background_Action}
&& S = \frac{1}{2 e^2}\mbox{tr}\,\mbox{Re}\int d^4x\,d^2\theta\,
\mbox{\boldmath$W$}^a \mbox{\boldmath$W$}_a - \frac{1}{64
e^2}\mbox{tr}\,\mbox{Re}\int d^4x\,d^4\theta\,\Bigg[16
\Big(e^{-2V}\mbox{\boldmath$D$}^a e^{2V}\Big)
\mbox{\boldmath$W$}_a
+\nonumber\\
&& + \Big(e^{-2V}\mbox{\boldmath$D$}^a e^{2V}\Big)
\bar{\mbox{\boldmath$D$}}^2 \Big(e^{-2V}\mbox{\boldmath$D$}_a
e^{2V}\Big) \Bigg] + \frac{1}{4}\int d^4x\, d^4\theta\,
\Big(\phi^+ e^{\mbox{\boldmath${\scriptstyle \Omega}$}^+} e^{2V}
e^{\mbox{\boldmath${\scriptstyle \Omega}$}} \phi +\tilde\phi^+
e^{-\mbox{\boldmath${\scriptstyle \Omega}$}^*} e^{-2V^t}
\times\nonumber\\
&& \times e^{-\mbox{\boldmath${\scriptstyle \Omega}$}^t}
\tilde\phi\Big) + \frac{1}{2}m\int d^4x\,
d^2\theta\,\tilde\phi^t\,\phi + \frac{1}{2}m\int d^4x\,
d^2\bar\theta\,\tilde\phi^+\phi^*,
\end{eqnarray}

\noindent where

\begin{equation}
\mbox{\boldmath$W$}_a = \frac{1}{8}
e^{\mbox{\boldmath${\scriptstyle \Omega}$}} \bar D^2
\Big(e^{-\mbox{\boldmath${\scriptstyle \Omega}$}}
e^{-\mbox{\boldmath${\scriptstyle \Omega}$}^+} D_a
e^{\mbox{\boldmath${\scriptstyle \Omega}$}^+}
e^{\mbox{\boldmath${\scriptstyle \Omega}$}} \Big)
e^{-\mbox{\boldmath${\scriptstyle \Omega}$}}.
\end{equation}

\noindent Action of the covariant derivatives on the field $V$ in
the adjoint representation is defined by the standard way.

It is convenient to choose a regularization and gauge fixing so
that invariance (\ref{Background_Transformations}) will be
unbroken. First, we fix a gauge by adding

\begin{equation}\label{Gauge_Fixing}
S_{gf} = - \frac{1}{32 e^2}\,\mbox{tr}\,\int d^4x\,d^4\theta\,
\Bigg(V \mbox{\boldmath$D$}^2 \bar{\mbox{\boldmath$D$}}^2  V + V
\bar {\mbox{\boldmath$D$}}^2 \mbox{\boldmath$D$}^2 V\Bigg)
\end{equation}

\noindent to the action. In this case terms quadratic in the
superfield $V$ will have the simplest form:

\begin{equation}
\frac{1}{2 e^2}\mbox{tr}\,\mbox{Re}\int d^4x\,d^4\theta\, V
\mbox{\boldmath$D$}_\mu^2 V.
\end{equation}

\noindent The corresponding action for the Faddeev--Popov ghosts
$S_{c}$ is written as

\begin{eqnarray}\label{Ghost_Action}
S_{c} = i\,\mbox{tr}\int d^4x\,d^4\theta\,\Bigg\{(\bar c + \bar
c^+) V \Big[(c + c^+) + \mbox{cth}\,V (c-c^+) \Big]\Bigg\},
\end{eqnarray}

\noindent The superfield $V$ in this expression is decomposed with
respect to the generators of the adjoint representation of a gauge
group, and the fields $c$ and $\bar c$ are the anticommuting
background chiral fields.

Moreover \cite{West}, the quantization procedure also requires
adding the action for the Nielsen--Kallosh ghosts

\begin{eqnarray}\label{B_Action}
S_B = \frac{1}{2e^2}\mbox{tr}\int d^4x\,d^4\theta\,B^+
e^{\mbox{\boldmath${\scriptstyle \Omega}$}^+}
e^{\mbox{\boldmath${\scriptstyle \Omega}$}}\,B,
\end{eqnarray}

\noindent where $B$ is an anticommuting chiral superfield, and the
background field should be decomposed with respect to the
generators of the adjoint representation of a gauge group. Because
the fields $B$ and $B^+$ do not interact with the quantum gauge
field, they contribute only to the one-loop (including
subtraction) diagrams. It is important to note that the factor
$1/e^2$ in action (\ref{B_Action}) is the same as in action for
the gauge fixing terms (\ref{Gauge_Fixing}).

The gauge fixing breaks the invariance of the action under quantum
gauge transformations (\ref{Quantum_Transformations}), but there
is a remaining invariance under BRST-transformations. The
BRST-invariance leads to Slavnov--Taylor identities, which relate
vertex functions of the quantum gauge field and ghosts.

For regularization we use the following method: Let us add the
term

\begin{eqnarray}\label{Regularized_Action}
&& S_{\Lambda} = \frac{1}{2 e^2}\mbox{tr}\,\mbox{Re}\int
d^4x\,d^4\theta\, V\frac{(\mbox{\boldmath$D$}_\mu^2)^{n+1}}{
\Lambda^{2n}} V
\end{eqnarray}

\noindent to action (\ref{Background_Action}). Then the invariance
under both supersymmetry transformations and transformations
(\ref{Background_Transformations}) is unbroken. Therefore, the
effective action, calculated with the background field method, is
invariant under both supersymmetry and background gauge
transformations. The described way of regularization is a bit
different from the method, proposed in Ref. \cite{West_Paper}. The
difference is in the form of the term, which contains higher
covariant derivatives. In the method, considered here, it breaks
the BRST-invariance, but the form of terms, quadratic in the
quantum superfield $V$, is simpler. This simplifies calculations
in a certain degree, while all particular features of the higher
derivative regularization are the same in the both cases. However,
because the higher derivative term breaks the BRST-invariance, it
is necessary to use a special subtraction scheme, which cancels
noninvariant terms and guarantees fulfilling the Slavnov--Taylor
identities in each order of the perturbation theory. Such a scheme
was proposed in Refs. \cite{Slavnov1,Slavnov2} and generalized to
the supersymmetric case in Refs. \cite{Slavnov3,Slavnov4}. With
the background field method such a scheme is simpler, because the
background gauge invariance guarantees, for example, the
transversality of the two-point Green function of the gauge field.
Nevertheless, additional subtractions should be made for Green
functions, containing the ghost fields.

Let us construct the generating functional as follows:

\begin{eqnarray}\label{Generating_Functional}
&& Z[J,\mbox{\boldmath$\Omega$}] = \int D\mu\,\exp\Big\{i S + i
S_\Lambda + i S_{gf} + i S_{gh} + i S_{S} + i S_{\phi_0}
+\nonumber\\
&& \qquad\qquad\qquad\qquad\qquad + i \int d^4x\,d^4\theta\,\Big(J
+ J[\mbox{\boldmath$\Omega$}] \Big)
\Big(V'[V,\mbox{\boldmath$\Omega$}] - {\bf V} \Big) \Big\},\qquad
\end{eqnarray}

\noindent where $S_{gf}$ is gauge fixing terms
(\ref{Gauge_Fixing}) and $S_{gh} = S_c + S_B$ is the corresponding
action for the Faddeev--Popov and Nielsen--Kallosh ghosts. In all
expressions the coupling constant $e$ should be substituted by the
bare coupling constant $e_0$. $S_S$ denotes terms with sources for
chiral superfields. In the extended form they are written as

\begin{equation}\label{Sources}
S_S = \int d^4x\,d^2\theta\, \Big(j^t\,\phi + \tilde
j^t\,\tilde\phi \Big) + \int d^4x\,d^2\bar\theta\, \Big(j^+\phi^*
+ \tilde j^+ \tilde\phi^*\Big).
\end{equation}

\noindent Moreover, in generating functional
(\ref{Generating_Functional}) we introduce the additional sources

\begin{eqnarray}\label{New_Sources}
&& S_{\phi_0} = \frac{1}{4}\int d^4x\,d^4\theta\,\Big(\phi_0^+
e^{\mbox{\boldmath${\scriptstyle \Omega}$}^+} e^{2V}
e^{\mbox{\boldmath${\scriptstyle \Omega}$}} \phi + \phi^+
e^{\mbox{\boldmath${\scriptstyle \Omega}$}^+} e^{2V}
e^{\mbox{\boldmath${\scriptstyle \Omega}$}}
\phi_0 +\nonumber\\
&& \qquad\qquad\qquad\qquad + \tilde\phi_0^+
e^{-\mbox{\boldmath${\scriptstyle \Omega}$}^*} e^{-2V^t}
e^{-\mbox{\boldmath${\scriptstyle \Omega}$}^t} \tilde\phi +
\tilde\phi^+ e^{-\mbox{\boldmath${\scriptstyle \Omega}$}^*}
e^{-2V^t} e^{-\mbox{\boldmath${\scriptstyle \Omega}$}^t}
\tilde\phi_0 \Big),\qquad
\end{eqnarray}

\noindent where $\phi_0$, $\phi_0^+$, $\tilde\phi_0$ and
$\tilde\phi_0^+$ are arbitrary scalar superfields. In principle,
it is not necessary to introduce the term $S_{\phi_0}$ in the
generating functional, but the presence of the parameters $\phi_0$
is highly desirable for investigating the Schwinger-Dyson
equations. The superfield ${\bf V}$ is defined by

\begin{equation}\label{Background Field}
e^{2{\bf V}} \equiv e^{\mbox{\boldmath${\scriptstyle \Omega}$}^+}
e^{\mbox{\boldmath${\scriptstyle \Omega}$}},
\end{equation}

\noindent and $J[\mbox{\boldmath$\Omega$}]$ is a so far undefined
functional. A reason of its introducing will be clear later. The
functional integration measure is written as

\begin{equation}
D\mu = DV\,D\bar c\,Dc\,DB\,D\phi\,D\tilde\phi.
\end{equation}

In order to understand how generating functional
(\ref{Generating_Functional}) is related with the ordinary
effective action, we perform the substitution $V \to V'$. Then we
obtain

\begin{equation}
Z[J,\mbox{\boldmath$\Omega$}] = \exp\Big\{ -i \int
d^4x\,d^4\theta\,\Big(J + J[\mbox{\boldmath$\Omega$}]\Big) {\bf V}
\Big\} Z_0\Big[J + J[\mbox{\boldmath$\Omega$}],
\mbox{\boldmath$\Omega$}\Big],
\end{equation}

\noindent where

\begin{equation}
Z_0[J,\mbox{\boldmath$\Omega$}] = \int D\mu\,\exp\Big\{i S + i
S_\Lambda + i S_{gf} + i S_{gh} + iS_S + iS_{\phi_0}+ i \int
d^4x\,d^4\theta\,J V \Big\}.
\end{equation}

\noindent If the dependence of $S$, $S_\Lambda$, $S_{gf}$,
$S_{gh}$, and $S_{\phi_0}$ on the arguments $V$,
$\mbox{\boldmath$\Omega$}$, and $\mbox{\boldmath$\Omega^+$}$ were
factorized into the dependence on the variable $V'$, $Z_0$ would
not depend on $\mbox{\boldmath$\Omega$}$ and
$\mbox{\boldmath$\Omega^+$}$ and would coincide with the ordinary
generating functional. This really takes place for action
(\ref{SYM_Action}). However, in the term with the higher
derivatives, in the gauge fixing terms, and in the ghost
Lagrangian such factorization does not occur. Therefore, $Z_0$
actually differs from the usual generating functional.

Using the functional $Z[J,\mbox{\boldmath$\Omega$},j]$ it is
possible to construct the generating functional for the connected
Green functions

\begin{eqnarray}
&& W[J,\mbox{\boldmath$\Omega$},j] = -i\ln
Z[J,\mbox{\boldmath$\Omega$},j]
=\nonumber\\
&&\qquad\qquad = - \int d^4x\,d^4\theta \Big(J +
J[\mbox{\boldmath$\Omega$}]\Big) {\bf V} + W_0\Big[J +
J[\mbox{\boldmath$\Omega$}],\mbox{\boldmath$\Omega$},j\Big]\qquad
\end{eqnarray}

\noindent and the corresponding effective action

\begin{eqnarray}\label{Effective_Action_Definition}
&& \Gamma[V,\mbox{\boldmath$\Omega$}, \phi] = -\int
d^4x\,d^4\theta\,\Big(J {\bf V} + J[\mbox{\boldmath$\Omega$}] {\bf
V}\Big) + W_0\Big[J +
J[\mbox{\boldmath$\Omega$}],\mbox{\boldmath$\Omega$},j\Big] -
\nonumber\\
&& - \int d^4x\,d^4\theta\,J V - \int d^4x\,d^2\theta\,
\Big(j^t\,\phi + \tilde j^t\,\tilde\phi \Big) - \int
d^4x\,d^2\bar\theta\, \Big(j^+ \phi^* + \tilde j^+ \tilde\phi^*
\Big).
\end{eqnarray}

\noindent The sources should be expressed in terms of fields using
the equations

\begin{eqnarray}
&& V = \frac{\delta}{\delta J} W[J,\mbox{\boldmath$\Omega$},j] = -
{\bf V} + \frac{\delta}{\delta J} W_0\Big[J +
J[\mbox{\boldmath$\Omega$}],\mbox{\boldmath$\Omega$},j\Big];
\nonumber\\
&& \phi = \frac{\delta}{\delta j} W[J,\mbox{\boldmath$\Omega$},j]
= \frac{\delta}{\delta j} W_0\Big[J +
J[\mbox{\boldmath$\Omega$}],\mbox{\boldmath$\Omega$},j\Big],\quad\mbox{e.t.c.}
\end{eqnarray}

\noindent Substituting these expressions into Eq.
(\ref{Effective_Action_Definition}), we write the effective action
as

\begin{eqnarray}
&& \Gamma[V,\mbox{\boldmath$\Omega$},\phi] = W_0\Big[J +
J[\mbox{\boldmath$\Omega$}],\mbox{\boldmath$\Omega$},j\Big] - \int
d^4x\,d^4\theta\,\Big(J[\mbox{\boldmath$\Omega$}] {\bf V} + J
\frac{\delta}{\delta J} W_0\Big[j,J +
J[\mbox{\boldmath$\Omega$}],\mbox{\boldmath$\Omega$},j\Big]\Big)
-\nonumber\\
&& - \int d^4x\,d^2\theta\,\phi\frac{\delta}{\delta j} W_0\Big[J +
J[\mbox{\boldmath$\Omega$}],\mbox{\boldmath$\Omega$},j\Big]
-\Big(\mbox{similar terms with $\phi^+$, $\tilde\phi$, and
$\tilde\phi^+$}\Big).
\end{eqnarray}

\noindent Let us now set $V = 0$, so that

\begin{equation}\label{Phi_To_Zero}
{\bf V} = \frac{\delta}{\delta J} W_0\Big[J +
J[\mbox{\boldmath$\Omega$}],\mbox{\boldmath$\Omega$},j\Big].
\end{equation}

We also take into account that the invariance under background
gauge transformations (\ref{Background_Transformations})
essentially restricts the form of the effective action. If the
quantum field $V$ in the effective action is set to 0, then the
superfield $K$ will be present only in the gauge transformation
law of the fields $\mbox{\boldmath$\Omega$}$ and
$\mbox{\boldmath$\Omega$}^+$, the only invariant combination being
expression (\ref{Background Field}). (It is invariant in a sense
that the corresponding transformation law does not contain the
superfield $K$.) This means that in the final expression for the
effective action we can set

\begin{equation}\label{K_Fixing}
\mbox{\boldmath$\Omega$} = \mbox{\boldmath$\Omega$}^+ = {\bf V}.
\end{equation}

\noindent In this case the effective action is

\begin{eqnarray}\label{Background_Gamma}
&& \Gamma[0,{\bf V},\phi] = W_0\Big[J + J[{\bf V}],{\bf V},j\Big]
- \int d^4x\,d^4\theta\,\Big(J + J[{\bf V}]\Big)
\frac{\delta}{\delta J} W_0\Big[J + J[{\bf V}], {\bf V}, j\Big]
-\nonumber\\
&& - \int d^4x\,d^2\theta\,\phi\frac{\delta}{\delta j} W_0\Big[J +
J[{\bf V}],{\bf V},j\Big] -\Big(\mbox{similar terms with $\phi^+$,
$\tilde\phi$, and $\tilde\phi^+$}\Big).
\end{eqnarray}

\noindent Note that this expression does not depend on form of the
functional $J[\mbox{\boldmath$\Omega$}]$. In particular, it can be
chosen to cancel terms linear in the field $V$ in Eq.
(\ref{Generating_Functional}). Such a choice will be very
convenient below.

If the gauge fixing terms and the terms with higher derivatives
depended only on $V'$, expression (\ref{Background_Gamma}) would
coincide with the ordinary effective action. However, as we
already mentioned above, the dependence on $V$,
$\mbox{\boldmath$\Omega$}$, and $\mbox{\boldmath$\Omega$}^+$ is
not factorized into the dependence on $V'$ with the proposed
method of regularization and gauge fixing. According to Ref.
\cite{Kluberg1,Kluberg2}, the invariant charge (and, therefore,
the Gell-Mann--Low function) is gauge independent, and the
dependence of the effective action on gauge can be eliminated by
renormalization of the wave functions of the gauge field, ghosts,
and matter fields. Therefore, for calculating the Gell-Mann-Low
function we may use the background gauge described above. We note
that if this gauge is used, the renormalization constant of the
gauge field $A_\mu$ is 1 due to the invariance of the action under
transformations (\ref{Background_Transformations}).

Nevertheless, generating functional (\ref{Generating_Functional})
is not yet completely constructed. The matter is that adding the
term with higher derivatives does not remove divergences from
one-loop diagrams. To regularize them, it is necessary to insert
the Pauli-Villars determinants in the generating functional
\cite{Slavnov_Book}. The Pauli-Villars fields should be introduced
for the quantum gauge field, ghosts (including the
Nielsen--Kallosh ghosts) and the matter superfields. Constructing
them we will at once use condition (\ref{K_Fixing}). So, we should
insert in the generating functional the factors

\begin{equation}\label{PV_Insersion}
\prod\limits_i \Big(\det PV(V,{\bf V},M_i)\Big)^{c_i},
\end{equation}

\noindent in which the Pauli-Villars determinants are defined by

\begin{equation}\label{PV_Determinants}
\Big(\det PV(V,{\bf V},M)\Big)^{-1} = \int DV_{PV}D\bar
c_{PV}Dc_{PV}DB_{PV}D\phi_{PV}D\tilde\phi_{PV}\exp(i S_{PV}),
\end{equation}

\noindent where the action for the Pauli-Villars fields is

\begin{eqnarray}
&& S_{PV} = \mbox{tr}\,\mbox{Re}\int d^4x\,d^4\theta\, V_{PV}
\Big[\frac{1}{2 e_0^2}\mbox{\boldmath$D$}_\mu^2\Big(1 +
\frac{\mbox{\boldmath$D$}_\mu^{2n}}{\Lambda^{2n}}\Big) -
\frac{1}{e_0^2} \mbox{\boldmath$W$}^a \mbox{\boldmath$D$}_a +
\frac{1}{e^2} M_{V}^2\Big]V_{PV} +\qquad\nonumber\\
&& + \frac{1}{4}\mbox{tr}\int d^4x\,d^4\theta (\bar c_{PV} + \bar
c_{PV}^+) V \Big[(c_{PV} + c_{PV}^+) + \mbox{cth}\,V
(c_{PV}-c_{PV}^+) \Big] +\nonumber\\
&& + \Bigg(\frac{1}{2}M_c\,\mbox{tr}\int d^4x\,d^2\theta\,\bar
c_{PV}\, c_{PV} + \mbox{h.c.}\Bigg) +\frac{1}{4e_0^2}\mbox{tr}\int
d^4x\,d^4\theta\,B_{PV}^+ e^{2{\bf V}} B_{PV} +\nonumber\\
&& + \mbox{tr}\Bigg(\frac{1}{2e^2}\int d^4x\,d^2\theta\,M_B
B_{PV}^2 + \mbox{h.c.}\Bigg) + \frac{1}{4}Z \int d^4x\,d^4\theta\,
\Big(\phi_{PV}^+ e^{\bf V} e^{2V} e^{\bf V}
\phi_{PV} +\nonumber\\
&& + \tilde\phi_{PV}^+ e^{-{\bf V}^t} e^{-2V^t} e^{-{\bf
V}^t}\tilde\phi_{PV} \Big) + \Bigg(\frac{1}{2}\int
d^4x\,d^2\theta\, M \tilde\phi_{PV}^t \phi_{PV} +
\mbox{h.c.}\Bigg),
\end{eqnarray}

\noindent and $Z$ is a renormalization constant for the matter
superfields. The Grassmanian parity of the Pauli--Villars fields
is opposite to the Grassmanian parity of usual fields,
corresponding to them. The coefficients $c_i$ in Eq.
(\ref{PV_Insersion}) satisfy the conditions

\begin{equation}
\sum\limits_i c_i = 1;\qquad \sum\limits_i c_i M_i^2 = 0.
\end{equation}

\noindent Below, we assume that $M_i = a_i\Lambda$, where $a_i$
are some constants. Inserting the Pauli-Villars determinants
allows cancelling the remaining divergences in all one-loop
diagrams, including diagrams containing counterterm insertions.
(This is guaranteed because the masses of the gauge field and
Nielsen--Kallosh ghosts are multiplied by the renormalized
coupling constants, and the other terms are multiplied by the bare
ones. This will be discussed later in more details.)

In this paper we will calculate the Gell-Mann--Low function, which
is determined by dependence of the two-point Green function on the
momentum in the limit $m\to 0$. That is why we consider the
massless case and write terms in the effective action,
corresponding to the renormalized two-point Green function, as

\begin{equation}\label{D_Definition}
\Gamma^{(2)}_V = - \frac{1}{16\pi} \mbox{tr}\int
\frac{d^4p}{(2\pi)^4}\,d^4\theta\,{\bf V}(-p)\,\partial^2\Pi_{1/2}
{\bf V}(p)\, d^{-1}(\alpha,\mu/p),
\end{equation}

\noindent where $\alpha$ is a renormalized coupling constant. The
Gell-Mann--Low function, denoted by $\beta(\alpha)$, is defined by

\begin{equation}\label{Gell-Mann-Low_Definition}
\beta\Big(d(\alpha,\mu/p)\Big) = \frac{\partial}{\partial \ln p}
d(\alpha,\mu/p).
\end{equation}

The Gell-Mann--Low function is scheme independent. To prove this,
we will use the following statement as a starting point: If we fix
a normalization point $\mu\ll\Lambda$ and impose in this point the
boundary condition for the renormalized two-point Green function
$d(p/\mu=1)$, then the two-point Green function is uniquely
determined and does not depend on both renormalization and
regularization. For example, if two different regularizations (or
renormalization schemes) are used, then

\begin{equation}\label{D_Equality}
d_1\Big(\alpha_1(\mu),\frac{p}{\mu}\Big) =
d_2\Big(\alpha_2(\mu),\frac{p}{\mu}\Big),
\end{equation}

\noindent where $\alpha_i(\mu)$ and $d_i$ are the renormalized
coupling constants at the scale $\mu$ and the renormalized
two-point Green functions, obtained in the first and in the second
regularization respectively. Setting $p=\mu$ in Eq.
(\ref{D_Equality}), it is possible to find the dependence
$\alpha_1(\alpha_2)$. Therefore, two different regularizations
differ in a finite renormalization of the coupling constant. We
note that such a renormalization can be gauge dependent and cause
the gauge dependence of the effective action divergent part.
However, the Gell-Mann--Low function, which we will calculate
below in this paper, does not depend on such a finite
renormalization, because (setting $x\equiv \ln p/\mu$)

\begin{equation}
\beta_1\Big(d_1(\alpha_1,x)\Big) = \frac{\partial}{\partial x}
d_1(\alpha_1,x) =\frac{\partial}{\partial x} d_2(\alpha_2,x) =
\beta_2\Big(d_2(\alpha_2,x)\Big)=
\beta_2\Big(d_1(\alpha_1,x)\Big).
\end{equation}

\noindent Therefore, the Gell-Mann--Low function is independent of
the regularization. In particular, a regularization can break the
BRST-invariance if the renormalized effective action is obtained
by subtractions, restoring the Slavnov--Taylor identities.

The anomalous dimension is defined similarly. First we consider
the two-point Green function for the matter superfield in the
limit $m\to 0$:

\begin{equation}\label{Renormalized_Gamma_2}
\Gamma^{(2)}_\phi = \frac{1}{4} \int
\frac{d^4p}{(2\pi)^4}\,d^4\theta\,\Big(\phi^+(-p,\theta)\,\phi(p,\theta)
+ \tilde\phi^+(-p,\theta)\,\tilde\phi(p,\theta) \Big) \,
ZG(\alpha,\mu/p),
\end{equation}

\noindent where $Z$ denotes the renormalization constant for the
matter superfield. Then the anomalous dimensions is defined by

\begin{equation}
\gamma\Big(d(\alpha,\mu/p)\Big) = -\frac{\partial}{\partial\ln p}
\ln ZG(\alpha,\mu/p).
\end{equation}

%%%%%%%%%%%%%%%%%%%%%%%%%%%%%%%%%%%%%%%%%%%%%%%%%%%%%%%%%%%%%%%%

\section{Calculation of quantum corrections with the higher
derivative regularization} \label{Section_Calculations}

\subsection{Supersymmetric electrodynamics}
\hspace{\parindent}\label{Subsection_Electrodynamics}

Calculation of quantum corrections with the higher derivative
regularization was made first for the $N=1$ supersymmetric
electrodynamics. The matter is that in the Abelian case the term
with higher derivatives is simpler. Really, the superfield $W_a$
is gauge invariant in the Abelian case. Therefore, the higher
derivative term contains usual derivatives, instead of covariant
derivatives. This essentially simplifies the Feynman rules. For
the considered model in the Abelian case they can be formulated as
follows:

1. External lines gives the factor

\begin{equation}
\prod\limits_{E} \int \frac{d^4p_{{}_{E_V}}}{(2\pi)^4}
V(p_{{}_{E_V}}) \int \frac{d^4p_{{}_{E_\phi}}}{(2\pi)^4}
\phi(p_{{}_{E_\phi}}) \cdot \ldots\cdot (2\pi)^4
\delta\Big(\sum\limits_{E} p_{{}_E}\Big),
\end{equation}

\noindent where the index $E$ numerates external momentums.

2. Each internal line of the superfield $V$ corresponds to the
propagator

\begin{equation}
\frac{8e^2}{(k^2+i0) \Big(1+(-1)^n k^{2n}/\Lambda^{2n}\Big)} \,
\delta^4(\theta_1-\theta_2).
\end{equation}

3. Each internal line $\phi-\phi^*$ or $\tilde\phi-\tilde\phi^*$
corresponds to the propagator

\begin{equation}
-\frac{1}{16 (k^2+i0)}\bar D^2 D^2 \delta^4(\theta_1-\theta_2).
\end{equation}

4. The Pauli--Villars fields are present only in closed loops.
Each internal line $\Phi-\Phi^*$ or $\tilde\Phi-\tilde\Phi^*$
corresponds to the propagator

\begin{equation}
- \frac{1}{16(k^2-M_i^2+i0)}\, \bar D^2 D^2
\delta^4(\theta_1-\theta_2),
\end{equation}

\noindent and each internal line $\Phi-\tilde\Phi$ or
$\Phi^*-\tilde\Phi^*$ corresponds to

\begin{equation}
\frac{M_i}{4(k^2-M_i^2+i0)}\,\bar D^2 \delta^4(\theta_1-\theta_2)
\quad\mbox{and}\quad \frac{M_i}{4(k^2-M_i^2+i0)}\, D^2
\delta^4(\theta_1-\theta_2)
\end{equation}

\noindent respectively. For each loop with the Pauli--Villars it
is necessary to add ${\displaystyle - \sum\limits_i c_i}$.

5. Each loop gives the integration with respect to the loop
momentum ${\displaystyle \int \frac{d^4k}{(2\pi)^4}}$.

6. Each vertex produces integration with respect to the
corresponding $\theta$: ${\displaystyle \int d^4\theta}$.

7. It is necessary to calculate a numerical coefficient for each
diagram.

We first present an expression for the two-loop contribution to
the effective action, corresponding to the two-point Green
function of the matter superfield. It was found in Ref.
\cite{tmf2} and will be needed later for calculating the
three-loop $\beta$-function. This contribution can be written in
form (\ref{Renormalized_Gamma_2}), where

\begin{equation}
G = 1 + G_1 + G_2 + O(\alpha^3)
\end{equation}

\noindent is a result of calculating the two-loop diagrams without
insertions of counterterms on matter lines. $G_1$ and $G_2$ denote
the following Euclidean integrals:

\begin{eqnarray}\label{Z1_Definition}
&& G_1 \equiv -\int \frac{d^4k}{(2\pi)^4}\,
\frac{2e^2}{k^2 (k+p)^2 (1+k^{2n}/\Lambda^{2n})};\\
\label{Z2_Definition} && G_2 \equiv \int
\frac{d^4k}{(2\pi)^4}\,\frac{d^4l}{(2\pi)^4}\, \frac{4 e^4}{k^2
l^2 (k+p)^2 (l+p)^2 (1+ k^{2n}/\Lambda^{2n}) (1+
l^{2n}/\Lambda^{2n})}
+\nonumber\\
&& + \int \frac{d^4k}{(2\pi)^4}\,\frac{d^4l}{(2\pi)^4}\, \frac{4
e^4}{\displaystyle k^2 l^2 (l+p)^2 (k+l+p)^2 (1+
k^{2n}/\Lambda^{2n}) (1+ l^{2n}/\Lambda^{2n})}
-\nonumber\\
&& - \int \frac{d^4k}{(2\pi)^4}\,\frac{d^4l}{(2\pi)^4}\, \frac{4
e^4 (k+l+2p)^2}{k^2 (k+p)^2 l^2 (l+p)^2 (k+l+p)^2 (1+
k^{2n}/\Lambda^{2n}) (1+ l^{2n}/\Lambda^{2n})}
+\quad\nonumber\\
&& + \int \frac{d^4k}{(2\pi)^4} \,\frac{4 e^4}{k^2 (k+p)^2
(1+k^{2n}/\Lambda^{2n})^2} \Bigg(\int \frac{d^4l}{(2\pi)^4}\,
\frac{1}{l^2 (k+l)^2}
-\\
&& - \sum\limits_i c_i \int
\frac{d^4l}{(2\pi)^4}\,\frac{1}{(l^2+M_i^2)((k+l)^2+M_i^2)} -
\frac{1}{8\pi^2}\Big(\ln\frac{\Lambda}{\mu}+b_1\Big)
\Big(1+k^{2n}/\Lambda^{2n}\Big) \Bigg).\qquad\nonumber
\end{eqnarray}

\noindent Therefore \cite{3LoopHEP}, the two-loop renormalization
constant for the matter superfield can be written as

\begin{eqnarray}\label{Constant_Z}
&& Z(e,\Lambda/\mu) = 1 + \frac{\alpha}{\pi}\, \Big(\ln
\frac{\Lambda}{\mu}+z_1\Big)
+\nonumber\\
&& \qquad\qquad + \frac{\alpha^2}{\pi^2}\,\Big(\ln^2
\frac{\Lambda}{\mu} + z_1 \ln\frac{\Lambda}{\mu}\Big) - \gamma_2\,
\alpha^2\, \Big(\ln \frac{\Lambda}{\mu} + z_2\Big)+
O(\alpha^3),\qquad
\end{eqnarray}

\noindent where $\gamma_2\alpha^2$ is the two-loop anomalous
dimension. Here we assume that in order to cancel one-loop
divergences, the counterterms

\begin{eqnarray}
&& \Delta S = - \frac{1}{16\pi^2}
\Big(\ln\frac{\Lambda}{\mu}+b_1\Big) \mbox{Re}\int
d^4x\,d^2\theta\,W_a C^{ab} \Big(1+
\frac{\partial^{2n}}{\Lambda^{2n}}\Big) W_b
+\nonumber\\
&&\qquad\qquad\qquad\qquad\qquad +
\frac{e^2}{16\pi^2}\Big(\ln\frac{\Lambda}{\mu}+z_1\Big) \int
d^4x\,d^4\theta\, \Big(\phi^* e^{2V}\phi +\tilde\phi^*
e^{-2V}\tilde\phi\Big)\qquad
\end{eqnarray}

\noindent are added to the action, where $b_1$, $z_1$, and $z_2$
are arbitrary finite constants. Choosing them, we fix a
subtraction scheme. It is important to note that $Z G$ is finite
by construction up to terms of the order $\alpha^3$.

Let us proceed now to the calculation of the Gell-Mann--Low
function. Diagrams, which define it, were calculated in Ref.
\cite{3LoopHEP}. It is necessary to remember that each internal
line of matter superfields can correspond to both the fields
$\phi$, $\tilde\phi$ and the Pauli--Villars fields. The total
three-loop contribution to the effective action can be presented
in form (\ref{D_Definition}), where the (sufficiently large)
expression for the function $d^{-1}$, obtained by explicit
calculating Feynman diagrams, is presented in Ref.
\cite{3LoopHEP}. As a check of the calculations we verify
cancellation of all noninvariant terms proportional to

\begin{equation}
\int \frac{d^4p}{(2\pi)^4}\,d^4\theta\,V(p,\theta)\,V(-p,\theta).
\end{equation}

In order to construct the Gell-Mann--Low function we consider the
expression

\begin{eqnarray}
\frac{d}{d\ln\Lambda} d^{-1}(\alpha_0,\Lambda/p)\Bigg|_{p=0} =
16\pi^2 \frac{d}{d\ln\Lambda} (A_1 + A_2 + A_3 + A_4)\Bigg|_{p=0},
\end{eqnarray}

\noindent where $A_1$ is a one-loop result; $A_2$ is a sum of
two-loop diagrams, three-loop diagrams with two loops of matter
superfields, and counterterm diagrams, arising due to the
renormalization of the coupling constant; $A_3$ is a sum of
three-loop diagrams with a single loop of matter superfields; and
$A_4$ is a sum of diagrams with counterterms insertions on lines
of matter superfields. Results, found in Ref. \cite{3LoopHEP}, can
be written as

\begin{eqnarray}
&& \frac{dA_1}{d\ln\Lambda}\Bigg|_{p=0} = \sum\limits_i c_i \int
\frac{d^4q}{(2\pi)^4} \frac{1}{q^2} \frac{d}{dq^2}
\frac{q^4}{(q^2+M_i^2)^2};\\
\vphantom{0}\nonumber\\
&& \frac{d A_2}{d\ln\Lambda}\Bigg|_{p=0} = -2e^2 \int
\frac{d^4q}{(2\pi)^4} \frac{1}{q^2} \frac{d}{dq^2}
\frac{d}{d\ln\Lambda} \int \frac{d^4k}{(2\pi)^4}
\frac{1}{k^2\Big(1 +
k^{2n}/\Lambda^{2n}\Big)^2}\Bigg(\frac{1}{(k+q)^2}
-\nonumber\\
&& - \sum\limits_{j} c_j \frac{q^4}{((k+q)^2+M_j^2)(q^2+M_j^2)^2}
\Bigg) \Bigg[\Big(1+k^{2n}/\Lambda^{2n}\Big)
\Bigg(1+\frac{e^2}{4\pi^2}\Big(\ln\frac{\Lambda}{\mu}+b_1\Big)\Bigg)
-\nonumber\\
&& - 2 e^2 \Bigg(\int \frac{d^4t}{(2\pi)^4}\,\frac{1}{t^2 (k+t)^2}
- \sum\limits_i c_i \int \frac{d^4t}{(2\pi)^4}\,
\frac{1}{(t^2+M_i^2) ((k+t)^2+M_i^2)} \Bigg)\Bigg];\\
\vphantom{0}\nonumber\\
&&
\frac{dA_3}{d\ln\Lambda} \Big|_{p=0} = \int
\frac{d^4q}{(2\pi)^4}\,\frac{1}{q^2}\frac{d}{dq^2}
\frac{d}{d\ln\Lambda}\int \frac{d^4k}{(2\pi)^4}
\frac{d^4l}{(2\pi)^4} \frac{4 e^4}{k^2
\Big(1+k^{2n}/\Lambda^{2n}\Big)\,
l^2\Big(1+l^{2n}/\Lambda^{2n}\Big)}
\times\nonumber\\
&& \times \Bigg\{\frac{1}{(q+k)^2}\Bigg[\frac{1}{2 (q+l)^2}  -
\frac{2 k^2}{(q+k+l)^2 (q+l)^2} + \frac{1}{(q+k+l)^2} \Bigg]
+\sum\limits_{i} c_i
\frac{q^4}{(q^2+M_i^2)^2}\times\nonumber\\
&&\times \frac{1}{((q+k)^2+M_i^2)}\Bigg[- \frac{1}{2
((q+l)^2+M_i^2) } + \frac{2 k^2}{((q+k+l)^2+M_i^2) ((q+l)^2+M_i^2)}
-\nonumber\\
&& - \frac{1}{((q+k+l)^2+M_i^2) } - \frac{2
M_i^2}{((q+k)^2+M_i^2)((q+k+l)^2+M_i^2)
} -\nonumber\\
&&  -\frac{2 M_i^2}{(q^2+M_i^2) ((q+l)^2+M_i^2) } -
\frac{2M_i^2}{((q+l)^2+M_i^2) ((q+k+l)^2+M_i^2)}\Bigg] \Bigg\};\\
\vphantom{0}\nonumber\\
&& A_4\Big|_{p=0} = \sum\limits_i c_i \int
\frac{d^4q}{(2\pi)^4}\,\frac{1}{q^2}\frac{d}{dq^2}\Bigg\{ -
\Bigg(\frac{\alpha}{\pi}\Big(\ln\frac{\Lambda}{\mu}+z_1\Big) +
\frac{\alpha^2}{\pi^2}\Big(\ln^2\frac{\Lambda}{\mu} + z_1
\ln\frac{\Lambda}{\mu}\Big)
-\nonumber\\
&& - \gamma_2\,\alpha^2 \Big(\ln\frac{\Lambda}{\mu}+z_2\Big)
\Bigg) \frac{q^4}{(q^2+M_i^2)^2} -
\frac{e^4}{16\pi^4}\Big(\ln\frac{\Lambda}{\mu}+z_1\Big)^2
\frac{(6q^4 + 3q^2 M_i^2 + M_i^4)M_i^2}{2(q^2+M_i^2)^3} +\nonumber\\
&& + \Bigg(\frac{\alpha}{\pi}\ln\frac{\Lambda}{\mu}+g_1\Bigg)\,
\frac{d}{d\ln\Lambda} \int\frac{d^4k}{(2\pi)^4}
\frac{e^2}{k^2(1+k^{2n}/\Lambda^{2n})} \frac{2q^4}{((k+q)^2+M_i^2)
(q^2+M_i^2)^2}\Bigg\}.\nonumber\\
\end{eqnarray}

Thus, we reveal an important feature of the quantum corrections
structure: all contributions to the Gell-Mann--Low function in
supersymmetric theories are integrals of total derivatives.
Really, in the four-dimensional spherical coordinates

\begin{equation}\label{Integral_Of_Total_Derivative}
\int \frac{d^4q}{(2\pi)^4}\frac{1}{q^2} \frac{d}{dq^2} f(q^2) =
\frac{1}{16\pi^2} \Big(f(q^2=\infty) - f(q^2=0)\Big).
\end{equation}

\noindent Using this equality it is possible to calculate all
integrals, presented above. The result is

\begin{eqnarray}\label{D_Detivative}
&& \frac{dA_1}{d\ln\Lambda}\Bigg|_{p=0} = \frac{1}{16\pi^2};
\nonumber\\
&& \frac{d}{d\ln\Lambda} (A_2 + A_3)\Bigg|_{p=0} =
\frac{e^2}{8\pi^2} \int \frac{d^4k}{(2\pi)^4}
\frac{d}{d\ln\Lambda} \frac{1}{k^4(1 + k^{2n}/\Lambda^{2n})^2}
\Bigg[\Big(1+k^{2n}/\Lambda^{2n}\Big)\times\nonumber\\
&& \times
\Bigg(1+\frac{e^2}{4\pi^2}\Big(\ln\frac{\Lambda}{\mu}+b_1\Big)\Bigg)
- 2 e^2 \Bigg(\int \frac{d^4t}{(2\pi)^4}\,\frac{1}{t^2 (k+t)^2}
-\nonumber\\
&&\qquad\qquad\qquad\qquad\qquad\qquad\quad - \sum\limits_i c_i
\int \frac{d^4t}{(2\pi)^4}\,
\frac{1}{(t^2+M_i^2) ((k+t)^2+M_i^2)} \Bigg)\Bigg]-\nonumber\\
&& -\frac{1}{4\pi^2} \int \frac{d^4k}{(2\pi)^4}
\frac{d^4l}{(2\pi)^4}\,\frac{d}{d\ln \Lambda}\frac{e^4}{(1+k^{
2n}/\Lambda^{2n})\,(1+l^{2n}/\Lambda^{2n})} \Bigg\{ \frac{1}{2 k^4
l^4} - \frac{1}{k^4 l^2 (k+l)^2} \Bigg\} =\nonumber\\
&& = \frac{1}{16\pi^2} \frac{d}{d\ln\Lambda} \Big(- G_1-
G_2+\frac{1}{2}G_1^2\Big)\Bigg|_{p=0} = - \frac{1}{16\pi^2} \ln
G.\qquad
\end{eqnarray}

Because the integrals, defining $A_4$, depend only on $\Lambda/p$,
taking the limit $\Lambda\to\infty$ is equivalent to taking the
limit $p\to 0$. Hence, taking into account that $\sum\limits
c_i=1$, we obtain

\begin{eqnarray}
&& A_4\Big|_{p=0} =\frac{1}{16\pi^2} \Bigg\{-
\Bigg(\frac{\alpha}{\pi}\Big(\ln\frac{\Lambda}{\mu}+z_1\Big) +
\frac{\alpha^2}{\pi^2}\Big(\ln^2\frac{\Lambda}{\mu} + z_1
\ln\frac{\Lambda}{\mu}\Big) - \gamma_2\,\alpha^2
\Big(\ln\frac{\Lambda}{\mu}+z_2\Big) \Bigg) +\nonumber\\
&& + \frac{\alpha^2}{2\pi^2}\Big(\ln\frac{\Lambda}{\mu}+z_1\Big)^2
\Bigg\} + O(\alpha^3) = - \frac{1}{16\pi^2} \ln Z + O(\alpha^3).
\end{eqnarray}

Collecting all contributions to the function $d^{-1}$, we obtain

\begin{equation}
d^{-1}(e,\mu/p) = \frac{1}{\alpha_0} + \ln\frac{\Lambda}{p} -
\ln(ZG) + \mbox{finite terms}.
\end{equation}

\noindent Therefore, there are divergences only in the one-loop
approximation. (This was first noted in Ref. \cite{hep}.) In order
to compensate them the bare coupling constant should be presented
in the form

\begin{eqnarray}
\frac{1}{\alpha_0} = \frac{1}{\alpha} -
\frac{1}{\pi}\ln\frac{\Lambda}{\mu}-b_1+O(\alpha^3).
\end{eqnarray}

\noindent Then the final result can be written as

\begin{equation}
d^{-1}(e,\mu/p) = \frac{1}{\alpha} + \ln\frac{\mu}{p} - \ln(ZG) +
O(\alpha^3) + \mbox{finite terms}.
\end{equation}

\noindent Using this expression it is possible to construct the
three-loop Gell-Mann--Low function according to Eq.
(\ref{Gell-Mann-Low_Definition}). In this case it coincides with
the expansion of the exact NSVZ $\beta$-function

\begin{equation}
\beta(\alpha) = \frac{\alpha^2}{\pi} \Big(1-\gamma(\alpha)\Big) +
O(\alpha^5).
\end{equation}

It is worth mentioning a difference between the results of
calculations, made with the higher covariant derivative
regularization and with the dimensional reduction. With the higher
derivative regularization divergences are only in the one-loop
approximation, while with the dimensional reduction they appear in
all loops. According to Ref. \cite{HD_And_DRED} the difference of
the results for divergences arises because with the dimensional
reduction a contribution of diagrams with counterterm insertions
is 0, while with the higher derivative regularization it is not 0.
In the three-loop approximation the situation is completely
similar. The results for the sum of diagrams with counterterms
insertions differ due to the mathematical inconsistency of the
dimensional reduction, which was first pointed in Ref.
\cite{Siegel2}. In particular, the straightforward application of
the dimensional reduction to calculating anomalies gives zero. Due
to the same reasons (see Ref. \cite{HD_And_DRED}) the sum of
diagrams with the counterterm insertions, defining the rescaling
anomaly, which was investigated in Ref. \cite{Arkani} in details,
is also 0 with the dimensional reduction:

\begin{equation}
\Big\langle \exp\Bigg(i (Z-1)\, \frac{1}{4} \int
d^4x\,d^4\theta\,\Big(\phi^* e^{2V}\phi + \tilde\phi^*
e^{-2V}\tilde\phi\Big)\Bigg)\Big\rangle = 1.
\end{equation}

\noindent With the higher derivative regularization we find

\begin{eqnarray}\label{Generalization_Of_Konishi_Anomaly}
&& \Big\langle \exp\Bigg(i (Z-1)\, \frac{1}{4} \int
d^4x\,d^4\theta\,\Big(\phi^* e^{2V}\phi + \tilde\phi^*
e^{-2V}\tilde\phi\Big)\Bigg)\Big\rangle
=\nonumber\\
&& = \exp\Bigg(-i\ln Z \frac{1}{16\pi^2} \mbox{Re}\int
d^4x\,d^2\theta\,W_a C^{ab} W_b +\mbox{finite terms} \Bigg).
\qquad
\end{eqnarray}

\noindent This result, obtained with the higher derivative
regularization, in a certain degree confirms speculations, made in
Ref. \cite{SV}. According to this paper, the Wilsonian action
$S_W$ is exhausted at the one-loop, while the effective action
$\Gamma$ has corrections in all loops. Now we see that $S_W$
should be replaced by the usual renormalized action.

It is important to note that the existence of divergences only in
the one-loop approximation {\it is not a physical result}. The
renormalized effective action and the Gell-Mann--Low function are
the same in both regularizations. The relation between the
Gell-Mann--Low function and the nonphysical $\beta$-function,
defined by divergence, is broken because the way of introducing
the regularization produces the dependence of the generating
functional on the normalization point $\mu$ at a fixed value
$e_0$.

\subsection{Yang--Mills theory without matter fields}
\label{Subsection_2Loop}
\hspace{\parindent}

In the previous section we demonstrated that in the $N=1$
supersymmetric electrodynamics all integrals, defining the
Gell-Mann--Low function in the limit $p\to 0$ were factorized into
integrals of total derivatives. Does this fact take place in the
non-Abelian case? To answer this question in Ref. \cite{2LoopYM}
the two-loop Gell-Mann--Low function was calculated for the pure
$N=1$ supersymmetric Yang--Mills theory (without matter fields)
with the higher covariant derivative regularization.

The one-loop $\beta$-function, calculated with the background
field method, is well-known \cite{West}. Using the higher
covariant derivative regularization does not essentially change
the calculation and its result \cite{PhysLett}. Let us mention the
typical features. The quantum superfield $V$ does not contribute
to the one-loop diagrams, because in the corresponding diagrams a
number of the spinor derivatives $D$, acting on propagators, is
less than 4. Really, a result of calculating any two-point diagram
is proportional to

\begin{equation}
\hat P_x \delta^8_{xy}\Big|_{x=y},
\end{equation}

\noindent where $x$ and $y$ are the points, to which the external
lines are attached. The result is not 0 only if the operator $\hat
P$ contains 4 spinor derivatives. However, two vertexes can
contain no more than 2 spinor derivatives, and propagators of the
gauge field do not contain spinor derivatives at all. Therefore,
all one-loop two-point diagrams are automatically 0. The one-loop
diagrams with the Pauli--Villars fields, corresponding to the
gauge field, are 0 due to the same reason. Since the higher
derivatives do not change a number of spinor derivatives in
vertexes, the one-loop contribution of the quantum field is also 0
in the regularized theory.

Therefore, the one-loop two-point Green-function of the gauge
field is completely determined by contributions of the
Faddeev--Popov and Nielsen--Kallosh ghosts. With the
regularization and gauge fixing, described above, the ghost
Lagrangians do not depend on the presence of higher derivative
terms. Due to anticommuting, the contributions of each ghost field
have opposite sign in comparison with the contribution of a chiral
scalar superfield in the adjoint representation of a gauge group.
Therefore, in the one-loop approximation the Gell-Mann--Low
function is

\begin{equation}
\beta(\alpha) = - \frac{3 C_2 \alpha^2}{2\pi} + O(\alpha^3).
\end{equation}

\begin{figure}[h]
\includegraphics[scale=0.75]{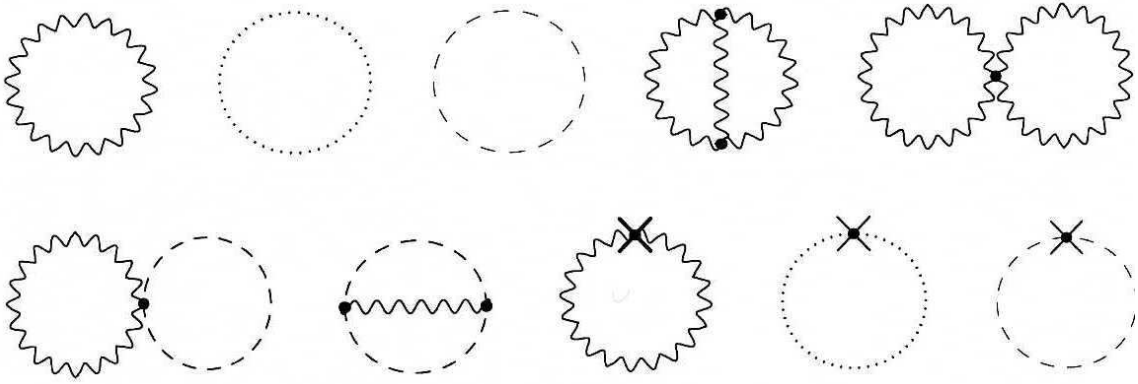}
\caption{Diagrams, contributing to the two-loop $\beta$-function
of the $N=1$ supersymmetric Yang--Mills theory.}
\label{Figure_Diagrams}
\end{figure}

The effective action in the two-loop approximation is calculated
by the standard way. It is contributed by diagrams, schematically
presented in Fig. \ref{Figure_Diagrams}. Usual diagrams are
obtained by attaching to them two external lines of the background
gauge field by all possible ways. In Fig. \ref{Figure_Diagrams} a
propagator of the quantum field $V$ is denoted by a wavy line, a
propagator of the Faddeev--Popov ghosts by dashes, and a
propagator of the Nielsen--Kallosh ghosts by dots. (We note that
they contribute only in the one-loop approximation, because the
Nielsen--Kallosh ghosts interact only with the background field.)

With the higher derivative regularization the propagator of the
quantum field is

\begin{equation}
\frac{1}{q^2 (1+q^{2n}/\Lambda^{2n})}
\end{equation}

\noindent (in the Euclidean space after the Weak rotation).
Feynman rules for vertexes, containing two lines of the quantum
field $V$, are also changed. In particular, the vertex with a
single line of the background superfield ${\bf V}$, which has the
momentum $p$, (it is denoted by a bold wavy line) is

\begin{equation}
\hspace*{-11cm}
\includegraphics[scale=0.5]{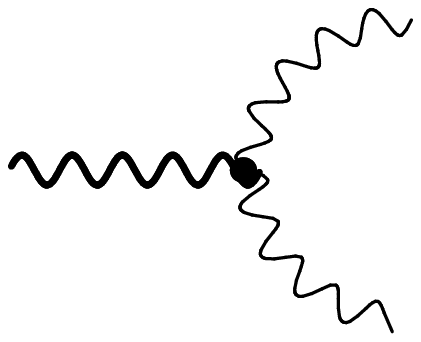}
\begin{picture}(-2,-0.7)(-2,-0.7)
\put(-2,0){${\displaystyle \sim \frac{1}{4}(2k+p)_\mu \bar D
\gamma^\mu\gamma_5 D {\bf V}
\Bigg(1+\frac{(k+p)^{2n+2}-k^{2n+2}}{\Lambda^{2n}\Big((k+p)^2-k^2\Big)}
\Bigg),}$}
\end{picture}
\end{equation}

\noindent and the vertex with two lines of the background
superfield ${\bf V}$, which have momentums $p$ and $-p$, is

\begin{eqnarray}
&& \hspace*{-0.7cm}\includegraphics[scale=0.6]{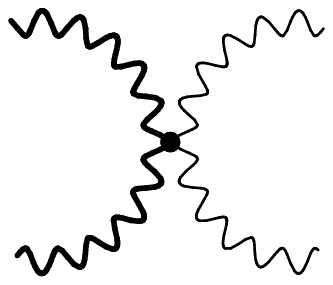}
\begin{picture}(-2,-0.7)(-2,-0.7)
\put(-2,0){${\displaystyle \sim \Big(4 {\bf V}
\partial^2\Pi_{1/2}{\bf V} +p^2{\bf
V}^2\Big)\Big(1+(n+1)\frac{k^{2n}}{\Lambda^{2n}}\Big)
+\frac{1}{\Lambda^{2n}}\Big((2k+p)^2{\bf V}
\partial^2\Pi_{1/2}{\bf V}+}$}
\end{picture}\nonumber\\
&&  + {\bf V}^2((k+p)^2-k^2)^2 \Big)
\Bigg(\frac{(k+p)^{2n+2}-k^{2n+2}}{((k+p)^2-k^2)^2} -
\frac{(n+1)k^{2n}}{(k+p)^2-k^2}\Bigg) -4{\bf
V}\partial^2\Pi_{1/2}{\bf V}.\qquad
\end{eqnarray}

According to the calculations, the two-loop contribution of the
Faddeev--Popov ghosts to the Gell-Mann--Low function is 0 that
agrees, for example, with Ref. \cite{Grisaru}. (Integrals,
defining the two-point Green function, appeared to be some finite
constants for the ghosts.)

As we already mentioned, the total two-loop contribution of the
two-point diagrams to the effective action can be presented in
form (\ref{D_Definition}) due to the Slavnov--Taylor identity. To
find the function $d^{-1}$ up to an unessential constant, we
differentiate it with respect to $\ln\Lambda$, and then set the
external momentum $p$ to 0. Later we will see that the result is a
finite constant $d_1$:

\begin{equation}
\frac{d}{d\ln\Lambda} d^{-1}(\alpha,\Lambda/p)\Bigg|_{p=0} = d_1.
\end{equation}

\noindent Therefore, the function $d^{-1}$ depends on the momentum
logarithmically

\begin{equation}
d^{-1}(\alpha,\Lambda/p) = d_1 \ln\frac{\Lambda}{p} +\mbox{const}.
\end{equation}

Calculating explicitly two-loop diagrams, presented in Fig.
\ref{Figure_Diagrams} (so far without diagrams with counterterm
insertions), differentiating the result with respect to
$\ln\Lambda$, and, then, setting $p=0$, we obtain (in the
Euclidean space, after the Weak rotation)

\begin{eqnarray}\label{D2}
&& d_1 = 8\pi\cdot 6\pi\,\alpha_0\, \int \frac{d^4k}{(2\pi)^4}
\frac{1}{k^2} \frac{d}{dk^2}\frac{d}{d\ln\Lambda} \int
\frac{d^4q}{(2\pi)^4}
 \Big(q^2(1+q^{2n}/\Lambda^{2n})\Big)^{-1}
\Bigg\{\Big((q+k)^2\times\nonumber\\
&& \times (1+(q+k)^{2n}/\Lambda^{2n})\Big)^{-1} \Bigg[2(n+1)
\Big(1+k^{2n}/\Lambda^{2n}\Big)^{-1} - 2n
\Big(1+k^{2n}/\Lambda^{2n}\Big)^{-2} \Bigg]\Bigg\}.
\end{eqnarray}

\noindent It is important to note that taking the limit $p\to 0$
is rather nontrivial, because the final result can contain
infrared divergent terms, proportional to $p$ or $p^2$, or terms,
proportional to $p$, but giving a finite contribution to $d_1$.
However, the calculation shows that all such terms are cancelled.
Moreover, the sum of diagrams appeared to be a total derivative
with respect to the module of the loop momentum, so that the
integral with respect to $d^4k$, which is contained in Eq.
(\ref{D2}), can be easily calculated by Eq.
(\ref{Integral_Of_Total_Derivative}). All substitutions at the
upper limit are 0 due to the higher derivative regularization, and
only the substitution at the lower limit is nonzero. Using
equations, presented above, we obtain

\begin{equation}
d_1 = - 6\alpha_0 \frac{d}{d\ln\Lambda} \int \frac{d^4q}{(2\pi)^4}
\Big(q^2(1+q^{2n}/\Lambda^{2n})\Big)^{-2}.
\end{equation}

\noindent This integral can be also easily calculated in the
four-dimensional spherical coordinates:

\begin{equation}
d_1 = \frac{12\alpha_0}{\pi} \int \frac{d^4q}{(2\pi)^4}
\frac{1}{q^4} q^2 \frac{d}{dq^2}(1+q^{2n}/\Lambda^{2n})^{-2} =
\frac{3\alpha_0}{4\pi^3}(1+q^{2n}/\Lambda^{2n})^{-2}\Bigg|_{0}^\infty
= - \frac{3\alpha_0}{4\pi^2}.
\end{equation}

\noindent (We note that the result does not depend on the
regularization parameter $n$.) Therefore, in the two-loop
approximation

\begin{equation}\label{Two_Loop_D}
d_0^{-1}(\alpha_0,\Lambda/p) = \frac{1}{\alpha_0} - \frac{3
C_2}{2\pi} \ln\frac{\Lambda}{p} - \frac{3 \alpha_0
C_2^2}{(2\pi)^2}\ln\frac{\Lambda}{p} + O(\alpha_0^2),
\end{equation}

\noindent where $d_0$ denotes the function $d$, calculated without
diagrams with counterterms insertions.

Therefore, the Gell-Mann--Low function, defined by Eq.
(\ref{Gell-Mann-Low_Definition}), in the two-loop approximation is

\begin{equation}
\beta(\alpha) = - \frac{3 C_2 \alpha^2}{2\pi} - \frac{3\alpha^3
C_2^2}{(2\pi)^2} + O(\alpha^4)
\end{equation}

\noindent and coincides with the expansion of the exact NSVZ
$\beta$-function in the considered order. We note that this result
does not depend on a possible finite constant in Eq.
(\ref{Two_Loop_D}).

For calculating quantum corrections it is also necessary to take
into account diagrams with counterterms insertions. Usually,
adding counterterms is equivalent to splitting the bare coupling
constant into the renormalized coupling constant and some infinite
additional term. However, using noninvariant regularizations (and,
in particular, the regularization, breaking the BRST-invariance,
which is used here), it is also necessary to add counterterms,
restoring the Slavnov--Taylor identities \cite{Slavnov1,Slavnov2}
in each order of the perturbation theory. In general, it is
necessary to analyze such counterterms. However, in the considered
case the situation is simpler. Really, the one-loop two-point
Green function for the Faddeev--Popov ghosts is finite and does
not depend on regularization. Interaction of ghosts with the
background field is fixed by the background gauge invariance,
which is unbroken with the considered regularization. Therefore,
additional counterterms do not contribute to subtraction diagrams,
containing a loop of the Faddeev--Popov ghosts, in the two-loop
approximation. Moreover, terms with the Faddeev--Popov ghosts do
not evidently depend on whether the bare or renormalized coupling
constant is in the gauge fixing action. Hence, their contributions
do not also depend on a way of splitting the bare coupling
constant into the renormalized one and counterterms.

Quantizing the theory we also write the bare coupling constant
$e_0$ in the gauge fixing terms. Therefore, a part of the action,
quadratic in the quantum field, is written as

\begin{equation}
\frac{1}{2 e_0^2}\mbox{tr}\,\mbox{Re}\int d^4x\,d^4\theta\, V
\Big[\mbox{\boldmath$D$}_\mu^2\Big(1 +
\frac{\mbox{\boldmath$D$}_\mu^{2n}}{\Lambda^{2n}}\Big) + 2 W^a
\mbox{\boldmath$D$}_a\Big]V.
\end{equation}

\noindent Breaking the invariance under the BRST-transformations
can lead to the necessity of adding counterterms proportional to

\begin{equation}
\mbox{tr}\int d^4x\,d^4\theta V \mbox{\boldmath$D$}_\mu^2 V.
\end{equation}

\noindent (If the background field is 0, this follows from Refs.
\cite{Slavnov3,Slavnov4}. Terms, containing the background field,
can be restored from the background gauge invariance.) But this
means that all one-loop diagrams, {\it including diagrams with
insertions of both the counterterms, appearing due to the
renormalization of the coupling constant, and the additional
counterterms}, with a loop of the quantum field $V$, are 0,
because they can contain no more than 2 spinor derivatives.

At last, let us consider diagrams, containing a loop of the
Nielsen--Kallosh ghosts. Since the Nielsen--Kallosh ghosts exist
only in the one-loop approximation, there are no additional
counterterms, caused by the noninvariance of the regularization
under the BRST-transformations, in these diagrams. However, the
contribution of the counterterm diagrams is essential due to the
renormalization of the coupling constant. Really, the coefficient
in the action for the Nielsen--Kallosh ghosts should be the same
as in the gauge fixing terms. Therefore, it must contain the bare
coupling constant:

\begin{equation}
\frac{1}{4 e_0^2} \mbox{tr}\int d^8x\,B^+ e^{2{\bf V}} B.
\end{equation}

\noindent To regularize diagrams with counterterm insertions and a
loop of Nielsen--Kallosh ghosts, the action for the corresponding
Pauli--Villars fields should be written as

\begin{equation}\label{PV_Action}
\mbox{tr}\int d^4x\,\Big(\frac{1}{4 e_0^2}\int d^4\theta\,
B_{PV}^+ e^{2{\bf V}} B_{PV} + \frac{M_B}{2e^2} \int d^2\theta
B_{PV}^2 + \frac{M_B}{2e^2} \int d^2\theta (B_{PV}^+)^2\Big).
\end{equation}

\noindent This expression also contains the bare coupling constant
$e_0$. $M_B$ is proportional to the regularization parameter
$\Lambda$. Really, let us present a bare coupling constant as

\begin{equation}
\frac{1}{e_0^2} = \frac{1}{e^2} Z_3,
\end{equation}

\noindent where $e$ is the renormalized coupling constant, and
$Z_3$ is the renormalization constant. Then, expanding the
Pauli--Villars determinant for the Nielsen--Kallosh ghosts in
powers of $Z_3-1$, we obtain terms, regularizing diagrams with
insertions of counterterms.

However, due to inserting this determinant the generating
functional starts to depend on the normalization point at a fixed
bare coupling constant $e_0$, because the renormalized coupling
constant $e$ depends on $\mu$.

In the Abelian case calculating divergences for the action,
similar to (\ref{PV_Action}), was made, for example, in Ref.
\cite{hep}. In the considered case it is also necessary to take
into account a factor $-C_2/2$, which appears because the
Nielsen--Kallosh ghosts are in the adjoint representation of a
gauge group and anticommute. (There is only one matter superfield
now, instead of 2 matter superfields in the Abelian case.)
Moreover, the renormalization constant of a matter field $Z$
should be substituted for the constant $Z_3$. Taking into account
these comments, the result of Ref. \cite{hep} can be formulated as
follows. Contribution of the counterterm diagrams for the
Nielsen--Kallosh ghosts to $1/\alpha$ can be written as

\begin{equation}
\frac{C_2}{2\pi} \ln Z_3.
\end{equation}

To find this contribution in the two-loop approximation, we note
that after the one-loop renormalization the renormalization
constant will be

\begin{equation}
Z_3 = 1 + \frac{3C_2 \alpha}{2\pi}\ln\frac{\Lambda}{\mu} +
O(\alpha^2).
\end{equation}

\noindent Therefore, the contribution of diagrams with counterterm
insertions in the two-loop approximation is written as

\begin{equation}
\frac{3 \alpha C_2^2}{(2\pi)^2} \ln\frac{\Lambda}{\mu}.
\end{equation}

\noindent This contribution exactly cancels the two-loop
divergence, so that after the one-loop renormalization

\begin{equation}
d^{-1}(\alpha,\mu/p) = \frac{1}{\alpha} - \frac{3 C_2}{2\pi}
\ln\frac{\mu}{p} - \frac{3 \alpha C_2^2}{(2\pi)^2}\ln\frac{\mu}{p}
+ O(\alpha^2).
\end{equation}

For an arbitrary order of the perturbation theory it is reasonable
to propose that the two-point Green function of the gauge field is
given by

\begin{equation}\label{Alpha}
\frac{1}{d(\alpha,\mu/p)} = \frac{1}{\alpha_0} - \frac{1}{2\pi}
C_2 \ln d(\alpha_0,\Lambda/p) + \frac{1}{2\pi} C_2 \ln
Z_3(\alpha,\Lambda/\mu) - \frac{3}{2\pi} C_2 \ln\frac{\Lambda}{p}.
\end{equation}

\noindent Really, it is easy to see that the exact NSVZ
$\beta$-function is obtained by differentiating this equality with
respect to $\ln p$, and the term, proportional to $\ln Z_3$ is
obtained from contributions of diagrams with counterterms
insertions. In the two-loop approximation this equation agrees
with (\ref{Two_Loop_D}), if the contribution of diagrams with
counterterm insertions is taken into account.

If Eq. (\ref{Alpha}) is true, then divergences exist only in the
one-loop approximation. Really, because

\begin{equation}
\frac{1}{d(\alpha,\mu/p)} =
\frac{1}{d(\alpha_0(\alpha,\Lambda/\mu),\Lambda/p)}
Z_3(\alpha,\Lambda/\mu)
\end{equation}

\noindent is finite, it is necessary to cancel only the one-loop
divergence. For this purpose the bare coupling constant is
presented as

\begin{equation}
\frac{1}{\alpha_0} = \frac{1}{\alpha} + \frac{3}{2\pi} C_2
\ln\frac{\Lambda}{\mu}.
\end{equation}

\noindent We note that presence of divergences only in the
one-loop approximation in this case does not mean that the
physical $\beta$-function has only the one-loop contribution.
Really, the physical $\beta$-function is a derivative of the
two-point Green function with respect to the logarithm of the
momentum if proper boundary conditions are imposed. Such function,
as we already saw, has corrections in all loops. A relation
between the divergences and the physical $\beta$-function is
broken due to the way of the regularization of diagrams with the
counterterm insertions, which leads to the dependence of the
generating functional on a normalization point at a fixed bare
coupling constant \cite{HD_And_DRED}. Thus, similar to the
electrodynamics, we obtain that the renormalized effective action
is exhausted at the one-loop, while the Gell-Mann--Low function
has corrections in orders of the perturbation theory. Note, that
this conclusion agrees with Ref. \cite{Mas}, in which the two-loop
$\beta$-function for the $N=1$ supersymmetric Yang--Mills theory
was calculated with the differential renormalization \cite{DiffR}.

So, if Eq. (\ref{Alpha}) is valid, the Gell-Mann--Low function
coincides with the exact NSVZ $\beta$-function, and divergences in
the two-point Green function exist only in the one-loop
approximation.

%%%%%%%%%%%%%%%%%%%%%%%%%%%%%%%%%%%%%%%%%%%%%%%%%%%%%%%%%%%%%%%%

\section{Schwinger--Dyson equations and Slavnov--Taylor identities}
\label{Section_SD}

\subsection{Schwinger--Dyson equations for the contribution of
matter superfields}
\hspace{\parindent}

The calculations, described in the previous section, reveal some
interesting features of the quantum correction structure, which
appear if the higher derivatives are used for regularization. In
order to partially explain these features, it is possible to use a
method \cite{SD,SDYM}, based on substituting solutions of the
Slavnov--Taylor identities into the Schwinger--Dyson equations.
Here we will discuss only structure of the matter superfields
contribution to the exact $\beta$-function. Contribution of
diagrams with loops of the gauge fields and ghosts is not so far
calculated using this approach.

In order to construct the Schwinger-Dyson equations for model
(\ref{SYM_Action}) it is necessary to split the action into three
parts: the action for the background field, the kinetic term for
quantum fields, which does not contain the background field, and
interaction, in which the other terms are included:

\begin{equation}
S = S({\bf V}) + S_2(V,\phi) + S_I(V,{\bf V},\phi).
\end{equation}

\noindent (Earlier we saw that terms of the first order in the
superfield $V$, which were obtained from the expansion of the
classical action, can be omitted.) So, generating functional
(\ref{Generating_Functional}) can be written as

\begin{eqnarray}
&& Z[J,j,{\bf V}] = \int d\mu\,\exp\Big(i S[{\bf V}] + i
S_2[V,\phi] + i S_I[V,{\bf V},\phi] + i S_S + i \int d^8x\,J V
\Big)
=\\
&& = \exp\Big(i S[{\bf V}]+iS_I\Big[\frac{1}{i}
\frac{\delta}{\delta J},\frac{1}{i} \frac{\delta}{\delta {\bf J}},
\frac{1}{i}\frac{\delta}{\delta j} \Big]\Big) \times\nonumber\\
&& \qquad\qquad\qquad\qquad \times \int d\mu\,\exp\Big(i
S_2[V,\phi] + i S_S + i \int d^8x\,J V + i \int d^8x\,{\bf J} {\bf
V}\Big)\Bigg|_{{\bf J}=0},\qquad\nonumber
\end{eqnarray}

\noindent where ${\displaystyle \int d^8x \equiv \int
d^4x\,d^4\theta_x}$. Let us differentiate this expression with
respect to the background field

\begin{eqnarray}
&& \frac{\delta}{\delta {\bf V}_x} Z[J,j,{\bf V}] = i \frac{\delta
S[{\bf V}]}{\delta {\bf V}}\,Z +
\exp\Big(iS_I\Big[\frac{1}{i}\frac{\delta}{\delta
J},\frac{1}{i}\frac{\delta}{\delta {\bf J}},
\frac{1}{i}\frac{\delta}{\delta j} \Big]\Big) i {\bf J_x}
\times\nonumber\\
&& \qquad\qquad\qquad\times \int d\mu\,\exp\Big(i S_2[V,\phi] + i
S_S + i \int d^8x\, J V + i \int d^8x\,{\bf J} {\bf
V}\Big)\Bigg|_{{\bf J}=0}.\qquad
\end{eqnarray}

\noindent Moving the current ${\bf J}_x$ to the left and dividing
the result to $Z$, we obtain

\begin{equation}
\frac{\delta}{\delta {\bf V}_x} W[J,j,{\bf V}] = \frac{\delta
S[{\bf V}]}{\delta {\bf V}_x} + \frac{\delta}{\delta {\bf V}_x}
S_I\Big[{\bf V}, V + \frac{1}{i}\frac{\delta}{\delta J},
\phi+\frac{1}{i}\frac{\delta}{\delta j}\Big].
\end{equation}

\noindent Because the background field ${\bf V}$ is a parameter of
the effective action, these equality can be equivalently written
as

\begin{equation}
\frac{\delta\Gamma}{\delta {\bf V}_x} = \frac{\delta S[{\bf
V}]}{\delta {\bf V}_x} + \Big\langle \frac{\delta}{\delta {\bf
V}_x} S_I\Big[{\bf V}, V, \phi\Big]\Big\rangle,
\end{equation}

\noindent where the angular brackets denote taking an expectation
value by the ordinary functional integration. Certainly, it is
necessary to set the field $V$ (an argument of the effective
action) to 0 in the final result.

Let us find a contribution to this expression given by the matter
superfields. The corresponding interaction terms are

\begin{equation}
S_I = \frac{1}{4} \int d^8x\,\phi^+ \Big( e^{\bf V} e^{2V} e^{\bf
V} -1\Big) \phi + \mbox{similar terms with }\tilde\phi.
\end{equation}

\noindent Differentiating $S_I$ with respect to the background
field, we obtain that the corresponding contribution to the
effective action is written in the form

\begin{eqnarray}\label{SD_SYM}
&& \frac{\delta\Gamma}{\delta {\bf V}_x^a} = \mbox{Terms without
matter} + \sum\limits_i c_i \frac{\delta}{\delta {\bf
V}_x^a}\Big\langle\ln \det PV\Big(V,{\bf V},M_i\Big)\Big\rangle
+ \qquad\nonumber\\
&& + \frac{1}{4} \frac{\delta}{\delta {\bf V}_x^a} \int
d^8x\,\Big\langle \phi_x^+  e^{{\bf V}_x} e^{2V_x} e^{{\bf V}_x}
\phi_x + \mbox{similar terms with }\tilde\phi \Big\rangle.\quad
\end{eqnarray}

\noindent We are interested in the two-point Green function of the
gauge field, corresponding to the expansion of the effective
action in powers of the background field up to the second order
terms. This function is evidently symmetric with respect to the
indexes, numerating generators of the gauge group. Then, using the
form of action for additional sources (\ref{New_Sources}), we
easily obtain

\begin{equation}
\frac{1}{4} \Big\langle \phi_x^+ T^a e^{{\bf V}_x} e^{2V_x}
e^{{\bf V}_x} \phi_x + \phi_x^+ e^{{\bf V}_x} e^{2V_x} e^{{\bf
V}_x} T^a \phi_x \Big\rangle = \frac{1}{i}\mbox{tr}\Bigg[ T^a
\Bigg(\frac{\delta^2 \Gamma}{\delta j_x^+ \delta\phi_{0x}^+} +
\frac{\delta^2\Gamma}{\delta j_x \delta\phi_{0x}} \Bigg) \Bigg].
\end{equation}

\noindent (Here the derivatives with respect to the sources must
be expressed in terms of fields.)

Therefore, using Eq. (\ref{SD_SYM}) and taking into account
similar terms with the fields $\tilde\phi$, the corresponding
contribution to the two-point Green function of the matter
superfield can be written as

\begin{equation}\label{SD1}
\frac{\delta\Gamma}{\delta {\bf V}^b_y \delta {\bf V}^a_x} =
\ldots + e\,\frac{\delta}{\delta {\bf V}^b_y} \mbox{tr} \Bigg[T^a
\frac{1}{i}\frac{\delta^2 \Gamma}{\delta j_x^+\,
\delta\phi_{0x}^+} - (T^a)^t \frac{1}{i}
\frac{\delta^2\Gamma}{\delta\tilde j_x^+\,\delta\tilde\phi_{0x}^+}
+ \mbox{h.c.}\Bigg],
\end{equation}

\noindent where dots denote contributions of the gauge fields,
ghosts, and also all possible Pauli-Villars fields. We note that
the calculation of the Pauli-Villars fields contributions are made
completely similar to the calculation of ordinary fields
contributions \cite{SD}, and details of this calculation are not
presented here. The result will be given below.

Performing differentiation in Eq. (\ref{SD1}), this equation can
be graphically presented as a sum of two effective diagrams

\begin{equation}\label{SD_Equation}
\begin{picture}(0,1.8)
\put(-4.9,0.6){$\Delta\Gamma^{(2)}_V =$} \put(0.8,0.6){+}
\hspace*{-3.5cm}
\includegraphics[scale=0.88]{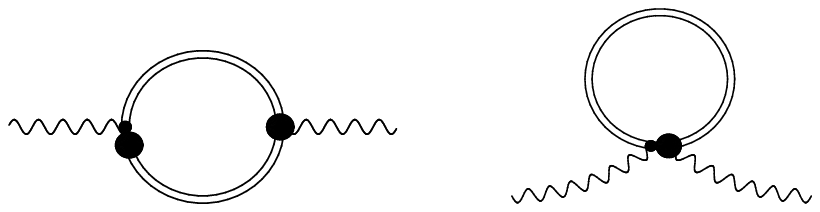}
\end{picture}
\end{equation}

\noindent The double lines correspond to the effective
propagators, which are written as

\begin{equation}\label{Inverse_Functions}
\Bigg(\frac{\delta^2\Gamma}{\delta\phi_x^+\delta\phi_y}\Bigg)^{-1}
= -  \frac{G D_x^2 \bar D_x^2}{4(\partial^2 G^2 + m^2 J^2)}
\delta^8_{xy};\qquad
\Bigg(\frac{\delta^2\Gamma}{\delta\phi_x\delta\tilde\phi_y}\Bigg)^{-1}
=  - \frac{m J \bar D_x^2}{\partial^2 G^2 + m^2 J^2}
\delta^8_{xy},
\end{equation}

\noindent depending on the chirality of the ends. The functions
$G$ and $J$ are determined by the two-point Green functions of the
matter superfield as

\begin{equation}\label{Explicit_Green_Functions}
\frac{\delta^2\Gamma}{\delta\phi_x^+\delta\phi_y} = \frac{D_x^2
\bar D_x^2}{16} G(\partial^2) \delta^8_{xy};\qquad
\frac{\delta^2\Gamma}{\delta\phi_x\delta\tilde\phi_y} = -
\frac{\bar D_x^2}{4} m J(\partial^2) \delta^8_{xy},
\end{equation}

\noindent where $\delta^8_{xy}\equiv
\delta^4(x-y)\delta^4(\theta_x-\theta_y)$.

\subsection{Slavnov--Taylor identities and their solutions}
\hspace{\parindent}

The vertex functions in Eq. (\ref{SD_Equation}) can be obtained
from the Slavnov-Taylor identities. Certainly, it is necessary to
take into account that we consider vertex functions, which have an
external line of the {\bf background} field. As we already
mentioned above, the effective action is invariant under the
background gauge transformations. It is easy to see that this
invariance can be expressed by the equality

\begin{eqnarray}\label{Invariance}
&& 0 = \int d^8y\,\Bigg(\frac{\delta\Gamma}{\delta {\bf
V}_y^a}\delta {\bf V}_y^a + \frac{\delta\Gamma}{\delta\phi_y}
\frac{D^2}{8\partial^2} \delta\phi_y + \delta\phi_y^+ \frac{\bar
D^2}{8\partial^2} \frac{\delta\Gamma}{\delta\phi_y^+} +
\frac{\delta\Gamma}{\delta\phi_{0y}} \delta\phi_{0y} +
\delta\phi_{0y}^+
\frac{\delta\Gamma}{\delta\phi_{0y}^+} +\qquad\nonumber\\
&& + \mbox{similar terms with $\tilde\phi$ and
$\tilde\phi_0$}\Bigg),
\end{eqnarray}

\noindent where

\begin{eqnarray}
&& \delta \phi_x = \Lambda_x \phi_x;\qquad \delta \phi_x^+ =
\phi_x^+ \Lambda_x^+;\qquad \delta \phi_{0x} = \Lambda_x
\phi_{0x};\qquad \delta \phi_{0x}^+ = \phi_{0x}^+ \Lambda_x^+;\nonumber\\
&& \delta {\bf V}_x = -\frac{1}{2}\Big(\Lambda_x +
\Lambda_x^+\Big) + O({\bf V}).
\end{eqnarray}

\noindent Here $\Lambda$ is an arbitrary chiral superfield, and
all other terms in $\delta {\bf V}$ are proportional at least to
the first degree of the background field. (For the fields
$\tilde\phi$ e.t.c. the transformation laws can be also easily
written.) Let us differentiate Eq. (\ref{Invariance}) with respect
to $\Lambda_y^a$, $\phi^+_{0z}$, $\phi_x$ or with respect to
$\Lambda_y^a$, $\tilde\phi_{0z}$, $\phi_x$. As a result we obtain
the Slavnov-Taylor identities

\begin{eqnarray}\label{ST_Identity}
&& 0 = \frac{1}{2e}(\bar D_y^2 + D_y^2) \frac{\delta\Gamma}{\delta
{\bf V}_y^a  \delta\phi^+_{0z}\delta\phi_x} - \frac{\delta\Gamma}{
\delta\phi_{0z}^+\delta\phi_y} T^a \bar D_y^2 \delta^8_{xy} -
D_y^2 \Big(\delta^8_{yz} T^a \frac{\delta\Gamma}{
\delta\phi_{0y}^+\delta\phi_x}\Big);\nonumber\\
&& 0 = \frac{1}{2e}(\bar D_y^2 + \bar D_y^2)
\frac{\delta^3\Gamma}{\delta {\bf V}^a_y \delta\tilde \phi_{0z}
\delta\phi_{x}} - \frac{\delta^2\Gamma}{\delta\tilde\phi_{0z}
\delta\phi_y} T^a \bar D_y^2 \delta^8_{xy} + \bar D_y^2 \Big(
\delta^8_{yz} T^a
\frac{\delta^2\Gamma}{\delta\tilde\phi_{0y}\delta\phi_x}\Big),\qquad
\end{eqnarray}

\noindent where

\begin{equation}\label{Useful_Identities}
\frac{\delta^2\Gamma}{\delta\phi_y \delta \phi_{0z}^+} = -
\frac{1}{8} G(\partial^2) \bar D_y^2\delta^8_{yz};\qquad
\frac{\delta^2\Gamma}{\delta\tilde\phi_y^* \delta \phi_{0z}^+} =
\frac{m}{32\partial^2} \Big(J(\partial^2)-1\Big) D_y^2 \bar D_y^2
\delta^8_{yz}.
\end{equation}

\noindent Solutions of Eqs. (\ref{ST_Identity}) are the functions

\begin{eqnarray}\label{Vertex3}
&& \frac{\delta^3\Gamma}{\delta {\bf
V}^a_y\delta\phi^+_{0z}\delta\phi_x}\Bigg|_{p=0} = e \Bigg[-2
\partial^2\Pi_{1/2}{}_y\Big(\bar D_y^2\delta^8_{xy}
\delta^8_{yz}\Big) F(q^2) + \frac{1}{8} D^b C_{bc} \bar
D_y^2\Big(\bar D_y^2\delta^8_{xy} D_y^c \delta^8_{yz} \Big) f(q^2)
-\vphantom{\frac{1}{2}}\nonumber\\
&& -\frac{1}{16} q^\mu G'(q^2) \bar D\gamma^\mu\gamma_5 D_y
\Big(\bar D_y^2\delta^8_{xy} \delta^8_{yz}\Big) -\frac{1}{4} \bar
D_y^2\delta^8_{xy} \delta^8_{yz}\, G(q^2)\Bigg] T^a;\\
\label{Vertex4} && \frac{\delta^3\Gamma}{\delta {\bf V}^a_y
\delta\tilde\phi_{0z} \delta\phi_x}\Bigg|_{p=0} = e
\Bigg[\,\frac{m}{32} D^b C_{bc} D_y^2 \Big(D_y^2 \bar
D_y^2\delta^8_{xy} D_y^c \delta^8_{yz}\Big) h(q^2) +\frac{m}{16}
J'(q^2) \Bigg( \bar D_y^2\delta^8_{xy} D_y^2 \delta^8_{yz}
-\nonumber\\
&& - D_y^2 \bar D_y^2 \delta^8_{xy} \delta^8_{yz} \Bigg)
+\frac{m}{16}\Bigg(\frac{J'(q^2)}{q^2} - \frac{J(q^2)-1}{q^4}
\Bigg) \Bigg(D_y^2 \bar D_y^2\delta^8_{xy} \frac{\bar D_y^2
D_y^2}{16} \delta^8_{yz} + D_y^2 \bar D_y^2 \delta^8_{xy} q^2
\delta^8_{yz} \Bigg)\Bigg] T^a.\nonumber\\
\end{eqnarray}

\noindent Here the primes denote derivatives with respect to
$q^2$,

\begin{equation}
\Pi_{1/2} = - \frac{1}{8 \partial^2} D^a \bar D^2 D_a = -
\frac{1}{8 \partial^2} \bar D^a D^2 \bar D_a
\end{equation}

\noindent is a supersymmetric transverse projection operator, and
the functions $F$, $f$ and $h$ can not be determined from the
Slavnov-Taylor identities.

Functions (\ref{Vertex3}) and (\ref{Vertex4}) allow finding
ordinary Green functions by the identities

\begin{equation}\label{Psi_Phi_Identites}
- \frac{D_z^2}{2}\frac{\delta^3\Gamma}{\delta {\bf
V}^a_x\delta\phi_y\delta\phi^+_{0z}} =
\frac{\delta^3\Gamma}{\delta {\bf
V}^a_x\delta\phi_y\delta\phi^+_z};\qquad - \frac{\bar D_z^2}{2}
\frac{\delta^3\Gamma}{\delta {\bf V}^a_x \delta\phi_y
\delta\tilde\phi_{0z}} = \frac{\delta^3\Gamma}{\delta {\bf V}^a_x
\delta\phi_y \delta\tilde\phi_z}.
\end{equation}

\noindent We note that the additional sources in Eq.
(\ref{New_Sources}) were specially introduced in order that such
identities take place. Using Eqs. (\ref{Psi_Phi_Identites}) we
find

\begin{eqnarray}\label{Vertex1}
&& \frac{\delta^3\Gamma}{\delta {\bf V}^a_y \delta\phi^+_z
\delta\phi_x} \Bigg|_{p=0} = e
\Bigg[\partial^2\Pi_{1/2}{}_y\Big(\bar D_y^2\delta^8_{xy} D_y^2
\delta^8_{yz}\Big) F(q^2) +\nonumber\\
&& \qquad\qquad +\frac{1}{32} q^\mu G'(q^2) \bar
D\gamma^\mu\gamma_5 D_y \Big(\bar D_y^2\delta^8_{xy} D_y^2
\delta^8_{yz}\Big) + \frac{1}{8} \bar D_y^2\delta^8_{xy} D_y^2
\delta^8_{yz}\, G(q^2)\Bigg] T^a;\\
\label{Vertex2} && \frac{\delta^3\Gamma}{\delta {\bf V}^a_y
\delta\tilde\phi_{z} \delta\phi_x }\Bigg|_{p=0} = -\frac{e m}{32}
J'(q^2) \Bigg[ \bar D_y^2\delta^8_{xy} D_y^2 \bar D_y^2
\delta^8_{yz} - D_y^2 \bar D_y^2 \delta^8_{xy} \bar
D_y^2\delta^8_{yz} \Bigg] T^a.\qquad
\end{eqnarray}

%%%%%%%%%%%%%%%%%%%%%%%%%%%%%%%%%%%%%%%%%%%%%%%%%%%%%%%%%%%%%%%%%%%%%%%%%

\subsection{Exact Gell-Mann--Low function.}
\label{Subsection_Two_Point_Function} \hspace{\parindent}

We will calculate

\begin{equation}\label{For_Calculation}
\frac{d}{d\ln\Lambda}\frac{\delta\Gamma}{\delta {\bf V}^b_y \delta
{\bf V}^a_x}\Bigg|_{p=0}.
\end{equation}

\noindent Note that the regularization by higher covariant
derivatives is essentially used here, because it allows
differentiating the integrand and taking the limit of zero
external momentum.

After substituting the vertex functions from Eqs. (\ref{Vertex1}),
(\ref{Vertex2}), (\ref{Vertex3}), (\ref{Vertex4}) and the
propagators from Eqs. (\ref{Inverse_Functions}), the Weak
rotation, and some simple transformations, using the algebra of
the covariant derivatives, we obtain that in the momentum
representation the first diagram is written as

\begin{eqnarray}\label{First_Diagram}
&& - C(R)\frac{d}{d\ln\Lambda} \mbox{tr}\int \frac{d^4p}{(2\pi)^4}
\frac{d^4q}{(2\pi)^4}\Bigg\{ {\bf V}\partial^2\Pi_{1/2}{\bf V}
\Bigg[\frac{8 G F}{q^2 G^2+m^2 J^2} - \frac{m^2 J J'}{2 q^2(q^2
G^2 + m^2 J^2)}  +\nonumber\\
&& + \frac{1}{2}\frac{d}{dq^2}\Bigg(\ln\Big(q^2 G^2 + m^2 J^2\Big)
+ \frac{m^2 J}{q^2 G^2 + m^2 J^2}\Bigg)\Bigg] + {\bf V}^2
\frac{G^2}{q^2 G^2 + m^2 J^2}\Bigg\},
\end{eqnarray}

\noindent and the second one is

\begin{eqnarray}\label{Second_Diagram}
&& - C(R)\frac{d}{d\ln\Lambda} \mbox{tr} \int
\frac{d^4p}{(2\pi)^4} \frac{d^4q}{(2\pi)^4} \Bigg\{{\bf
V}\partial^2\Pi_{1/2} {\bf V} \Bigg[-\frac{8 G F}{q^2 G^2+m^2 J^2}
- \frac{m^2 JJ'}{2 q^2(q^2G^2+m^2J^2)}
+\nonumber\\
&& + \frac{m^2 J (J-1)}{2 q^4 (q^2 G^2+m^2 J^2)} - \frac{8 G f + 8
m^2 J h}{q^2 G^2 + m^2 J^2} \Bigg] - {\bf V}^2 \frac{G^2}{q^2 G^2
+ m^2 J^2}\Bigg\}.
\end{eqnarray}

\noindent Let $d_0$ denotes the function $d$, calculated without
taking into account counterterm insertions on lines of matter
superfields, or, equivalently, at $Z=1$. Moreover, we take into
account that for finding the Gell-Mann--Low function it is
necessary to set $m=0$. Then, adding the results for the effective
diagrams in the Schwinger--Dyson equation and using Eq.
(\ref{D_Definition}), we obtain

\begin{eqnarray}\label{Result}
&& -\frac{1}{16\pi} \frac{d}{d\ln\Lambda} d_0^{-1}\Big|_{p=0} =
\ldots - C(R)\int\frac{d^4q}{(2\pi)^4}\frac{d}{d\ln\Lambda}
\Bigg\{ \frac{1}{2q^2}\frac{d}{dq^2} \ln\Big(q^2 G^2\Big) -
\frac{8 f}{q^2
G} -\qquad \nonumber\\
&& - \sum\limits_i c_i \frac{1}{2q^2}\frac{d}{dq^2}
\Bigg(\ln\Big(q^2 G_{PV}^2 + M_i^2 J_{PV}^2\Big) + \frac{M_i^2
J_{PV}}{q^2 G_{PV}^2 + M_i^2 J_{PV}^2}\Bigg) +\nonumber\\
&& +\sum\limits_i c_i \Bigg(\frac{M_i^2 J_{PV}\,\Big(2 q^2 J_{PV}'
- (J_{PV}-1)\Big)}{2 q^4\Big(q^2 G_{PV}^2 + M_i^2 J_{PV}^2\Big)} +
\frac{8 G_{PV} f_{PV} + 8 M_i^2 J_{PV} h_{PV}}{q^2 G_{PV}^2 +
M_i^2 J_{PV}^2}\Bigg)\Bigg\}.\qquad
\end{eqnarray}

\noindent (The dots denote contributions of the gauge fields and
ghosts, and the symbol $PV$ means that the corresponding function
is calculated for the Pauli--Villars fields.) We see that all
noninvariant terms, proportional to ${\bf V}^2$, and terms,
containing the unknown function $F$, are completely cancelled.
Nevertheless, there are the functions $f$ and $h$, which can not
be found from the Ward identities, in the final result.

It is important to note that many contributions in this expression
are integrals of total derivatives. This partially explains the
feature, noted earlier. However, calculations show that {\bf all}
terms in this expression should be integrals of total derivatives.
Moreover, the accurate analysis of the calculation, described
above, shows that terms, which are not factorized into total
derivatives in Eq. (\ref{Result}), are always equal to 0.
Therefore, it is possible to suggest the equality

\begin{equation}\label{New_Identity1}
\int\frac{d^4q}{(2\pi)^4} \frac{d}{d\ln\Lambda} \Bigg(\frac{m^2
J\,(2 q^2 J' - (J-1))}{2 q^4(q^2 G^2 + m^2 J^2)} + \frac{8 G f + 8
m^2 J h}{q^2 G^2 + m^2 J^2}\Bigg) = 0.
\end{equation}

\noindent For the massless theory it can be written in the simpler
form

\begin{equation}\label{Massless_New_Identity1}
\int\frac{d^4q}{(2\pi)^4} \frac{d}{d\ln\Lambda} \frac{f}{q^2 G} =
0.
\end{equation}

\noindent We will call Eqs. (\ref{New_Identity1}) and
(\ref{Massless_New_Identity1}) the new identity for the Green
functions. It is not a consequence of the gauge symmetry, the
supersymmetry, or the superconformal symmetry. Below we will
discuss it in more details, and now let us consider its
consequences. If the new identity for the Green functions is true,
then we obtain

\begin{eqnarray}\label{Effective_Action}
&& \frac{d}{d\ln \Lambda}\,
d_0^{-1}(\alpha_0,\Lambda/p)\Bigg|_{p=0} = 16\pi
C(R)\,\int\frac{d^4q}{(2\pi)^4} \frac{d}{d\ln\Lambda}
\frac{1}{2q^2} \frac{d}{dq^2}
\Bigg\{\ln(q^2 G^2) - \qquad\nonumber\\
&& - \sum\limits_i c_i \ln\Big(q^2 G_{PV}^2 + M_i^2 J_{PV}^2\Big)
-\sum\limits_i c_i \frac{M_i^2 J_{PV}}{q^2 G_{PV}^2 + M_i^2
J_{PV}^2} \Bigg\}.
\end{eqnarray}

The obtained integral is reduced to the total derivative in the
four-dimensional spherical coordinates, only the substitution at
the low limit being different from 0 \cite{SD}:

\begin{eqnarray}\label{Equation_For_D0}
&& \frac{d}{d\ln \Lambda}\,d_0^{-1}\Bigg|_{p=0} = -
\frac{C(R)}{2\pi}\,\frac{d}{d\ln\Lambda} \Bigg\{\ln G(0)^2 -
\sum\limits_i c_i \ln\Big(M_i^2 J_{PV}(0,a_i)^2\Big) -
\qquad\nonumber\\
&& -\sum\limits_i c_i \frac{1}{J_{PV}(0,a_i)} \Bigg\} =
\frac{1}{\pi} C(R) \Bigg(1-\frac{d\ln G}{d\ln \Lambda}
\Bigg)\Bigg|_{q=0}.
\end{eqnarray}

\noindent Here we took into account that the function
$J_{PV}(q^2/\Lambda^2,a_i)$ in the limit $q\to 0$ tended to some
finite constants, because it was defined by convergent (even in
the limit $\Lambda\to\infty$), dimensionless integrals, which did
not contain infrared divergences. Because both parts of Eq.
(\ref{Equation_For_D0}) depend on $\alpha_0$ and $\Lambda/p$, it
allows finding the expression for $d_0^{-1}$ up to an
insignificant numerical constant

\begin{equation}
d_0^{-1}(\alpha_0,\Lambda/p) = \frac{1}{\pi} C(R) \Big(\ln
\frac{\Lambda}{p}- \ln G(\alpha_0,\Lambda/p) \Big) + \mbox{const}.
\end{equation}

Now let us calculate the sum of diagrams with counterterm
insertions on lines of matter superfields exactly to all orders of
the perturbation theory. For this purpose we make the substitution

\begin{equation}
\phi \to \phi/\sqrt{Z};\quad \tilde\phi \to
\tilde\phi/\sqrt{Z};\quad \Phi \to \Phi/\sqrt{Z};\quad \tilde\Phi
\to \tilde\Phi/\sqrt{Z}
\end{equation}

\noindent in the generating functional. Then, it is easy to see
that if $\phi_0$ and $j$ are 0, the dependence on the
renormalization constant $Z$ can be found by the replacement

\begin{equation}\label{M_Substitution}
m \to m/Z;\qquad M_i \to M_i/Z.
\end{equation}

\noindent Making this replacement in Eq. (\ref{Equation_For_D0})
it is possible to restore the dependence of the effective action
on $Z$:

\begin{eqnarray}\label{Inverse_D}
&& d^{-1}(\alpha,\mu/p) = d_0^{-1}(\alpha_0,\Lambda/p) -
\frac{C(R)}{\pi}\ln Z(\alpha,\Lambda/\mu) =\nonumber\\
&&\qquad\qquad\qquad = \frac{1}{\alpha_0} + \frac{1}{\pi} C(R)
\Big(\ln \frac{\Lambda}{p}- \ln ZG(\alpha,\mu/p) \Big) +
\mbox{const}.\qquad
\end{eqnarray}

\noindent Differentiating this expression with respect to the
momentum $p$, we obtain the Gell-Mann--Low function

\begin{eqnarray}\label{Massless_D}
&& \beta(d) = \frac{\partial d}{\partial\ln p} = - d^2
\frac{\partial}{\partial\ln p}\,d^{-1} = \frac{d^2}{\pi} C(R)
\Big(1-\gamma(d)\Big).
\end{eqnarray}

\noindent This expression corresponds to a contribution of the
matter superfields to the exact NSVZ $\beta$-function
(\ref{NSVZ_Beta}). (It is necessary to take into account that in
the considered case the matter fields are in the representation
$R+\bar R$.) According to Eq. (\ref{Inverse_D}), to cancel the
dependence on $\Lambda$, the bare coupling constant should be
presented in the form

\begin{equation}
\frac{1}{\alpha_0} = \frac{1}{\alpha} - \frac{1}{\pi} C(R)
\ln\frac{\Lambda}{\mu}.
\end{equation}

\noindent This means that there are divergences only in the
one-loop approximation.

Note that if the new identity were not valid \cite{SDYM}, the
contribution of the matter superfields to the exact
$\beta$-function should be modified as follows:

\begin{equation}\label{Matter_Contribution}
\frac{\alpha^2 C(R)}{\pi}\Big(1-\gamma(\alpha) - \lim\limits_{p\to
0}\frac{16 p^2 f}{G}\Big).
\end{equation}

\section{New identity for Green functions}
\hspace{\parindent}\label{Section_New_Identity}

The new identity for the Green functions in the massive and
massless cases are given by Eqs. (\ref{New_Identity1}) and
(\ref{Massless_New_Identity1}) respectively. In the both cases it
can be equivalently rewritten in the following functional form
\cite{SDYM} (for the simplicity we will consider here the Abelian
case):

\begin{equation}\label{New_Identity_Functional_Form}
\int d^8x\,d^8y\,{\bf V}_y  D^a {\bf V}_x \frac{d}{d\ln\Lambda}
\frac{D_{az} \bar D_z^2}{\partial^2} \frac{\delta^3\Gamma}{\delta
j_z^+ \delta {\bf V}_y \delta \phi_{0x}^+}\Bigg|_{z=x, p=0} = 0,
\end{equation}

\noindent where we assume that the derivatives with respect to
sources should be expressed through the derivatives with respect
to fields. The condition $p=0$ means that in this case the
background field ${\bf V}$ depends only on variables $\theta$ and
is independent of the usual coordinates $x^\mu$. The derivative
with respect to $\ln\Lambda$ in this expression is essential. In
order to check this, let us consider for simplicity the massless
case and suppose that these derivative is absent. Then, from the
dimensional considerations

\begin{equation}\label{Integral}
\int \frac{d^4q}{(2\pi)^4} \frac{f}{q^2 G} = \int
\frac{d^4q}{(2\pi)^4} \frac{a(q^2/\Lambda^2)}{q^4},
\end{equation}

\noindent where $a$ is a dimensionless function, which is rapidly
decreasing at $q\to\infty$. In general, it is possible that
$a(0)\ne 0$. But if $a(0)\ne 0$, then the integral in Eq.
(\ref{Integral}) is not well defined: it is divergent in the
infrared region. In order to avoid this we introduce the
additional differentiation with respect to $\ln\Lambda$. Due to
its presence the term $a(0)$, which does not depend on $\Lambda$,
disappears, and the integral becomes finite in the infrared
region. Thus, without the derivative with respect to $\ln\Lambda$
the left hand side of the new identity is not well defined. It is
essential that the differentiation with respect to $\ln\Lambda$ is
possible due to using the higher covariant derivative
regularization.

To prove equality (\ref{New_Identity_Functional_Form}) it is
necessary to replace differentiation with respect to the sources
by the differentiation with respect to fields and substitute
expressions for the vertexes and propagators, presented earlier.

Let us rewrite identity (\ref{New_Identity_Functional_Form}) in a
different form. We commute the differentiations with respect to
${\bf V}$ and $\phi_0$, and again use the Schwinger--Dyson
equations, keeping only contributions of the matter superfields as
earlier. The result is

\begin{eqnarray}
&& \int d^8x\,d^8y\,{\bf V}_y D^a {\bf V}_x \frac{d}{d\ln\Lambda}
\frac{D_{az} \bar D_z^2}{\partial^2} \frac{\delta^2}{\delta j_z^+
\delta \phi_{0x}^+}\Big\langle(\phi^+ e^{2V}\phi - \tilde\phi^+
e^{-2V^t}\tilde\phi +\nonumber\\
&& + \phi_0^+ e^{2V}\phi - \tilde\phi_0^+ e^{-2V^t}\tilde\phi)_y
\Big\rangle\Bigg|_{z=x, p=0} = 0.
\end{eqnarray}

\noindent A contribution of terms, containing $\phi_0^+$ is 0,
because after performing the differentiation we obtain

\begin{equation}
\Big\langle \frac{D_{a} \bar D^2}{\partial^2} \phi_x^+ e^{2V_x}
\phi_x \Big\rangle = \frac{D_{az} \bar D_z^2}{\partial^2}
\frac{\delta^2\Gamma}{\delta j_z^+ \delta\phi_{0x}^+}\Bigg|_{z=x}
= 0.
\end{equation}

\noindent (In order to verify the last equality it is necessary to
express the derivative with respect to the source through
derivatives with respect to fields, use Eqs.
(\ref{Useful_Identities}) and (\ref{Inverse_Functions}), and take
into account that the expression $\hat P_x
\delta^4(\theta_x-\theta_y) \Big|_{\theta_x=\theta_y}$ is not 0
only if the operator $\hat P$ contains 4 spinor derivatives.

Therefore, the new identity can be written in terms of composite
operators correlators as

\begin{equation}
0 = \frac{d}{d\ln\Lambda} \Big\langle \int d^8x\,D^a {\bf
V}_x\Big(\frac{D_{a} \bar D^2}{\partial^2}\phi^+ e^{2V}
\phi\Big)_x \int d^8y\,{\bf V}_y \Big(\phi^+ e^{2V}\phi -
\tilde\phi^+ e^{-2V^t}\tilde\phi\Big)_y \Big\rangle\Big|_{p=0}.
\end{equation}

\noindent Taking into account the identity

\begin{equation}
\frac{D^2 \bar D^2}{\partial^2} \phi^+ = -16\phi^+
\end{equation}

\noindent this expression can be rewritten as

\begin{eqnarray}
&& 0 = \frac{d}{d\ln\Lambda} \Big\langle \int d^8x\,D^a {\bf
V}_x\Big(\frac{D_{a} \bar D^2}{\partial^2}\phi^+ e^{2V}
\phi\Big)_x\times\nonumber\\
&& \qquad\qquad\qquad \int d^8y\,{\bf V}_y \Big(\frac{D^2\bar
D^2}{\partial^2}\phi^+ e^{2V}\phi - \frac{D^2\bar D^2}{\partial^2}
\tilde\phi^+ e^{-2V^t}\tilde\phi\Big)_y
\Big\rangle\Big|_{p=0}.\qquad
\end{eqnarray}

\noindent Using the Leibnitz rule for the supersymmetric covariant
derivative, we find

\begin{equation}\label{Leibnitz_Rule}
\frac{D^2\bar D^2}{\partial^2}({\bf V}\phi^+) = {\bf V}
\frac{D^2\bar D^2}{\partial^2}\phi^+ + 2 D^a {\bf V} \frac{D_a\bar
D^2}{\partial^2}\phi^+ - 8i (\gamma^\mu C)_{ab} D^b \bar D^a {\bf
V} \frac{\partial_\mu}{\partial^2}\phi^+ +\ldots,
\end{equation}

\noindent where dots denote all terms, which do not contribute to
the considered correlator, because the momentum of the field ${\bf
V}$ is 0. (As a consequence, all terms, in which more than 4
spinor derivatives act on the fields ${\bf V}$ are 0.) The first
term in Eq. (\ref{Leibnitz_Rule}) also does not contribute to the
considered correlator, because (the sources are not set to 0)

\begin{eqnarray}
&& \frac{d}{d\ln\Lambda}\Big\langle\int d^8x\,\frac{D^2\bar
D^2}{16\partial^2}({\bf V}\phi^+) e^{2V} \phi\Big\rangle =
\frac{d}{d\ln\Lambda}\Big\langle\int d^8x\,({\bf V}\phi^+)
\frac{D^2\bar D^2}{16\partial^2}(e^{2V} \phi)\Big\rangle
=\qquad\nonumber\\
&& = \frac{d}{d\ln\Lambda}\int d^8x\,\frac{\delta}{\delta
j_x^+}\frac{D^2\bar D^2}{4\partial^2} \frac{\delta
\Gamma}{\delta\phi_{0x}^+} = -\frac{d}{d\ln\Lambda} \int
d^8x\,\frac{\delta}{\delta j_x^+}\frac{D^2}{2\partial^2}
\frac{\delta\Gamma}{\delta\phi_x^+} = 0.
\end{eqnarray}

\noindent Therefore, the new identity can be equivalently written
as

\begin{eqnarray}\label{New_Identity_Correlators}
&& \frac{d}{d\ln\Lambda} \Big\langle \int d^8x\,D^a {\bf V}_x
\frac{D_{a} \bar D^2}{\partial^2}\phi_x^+ e^{2V_x} \phi_x \int
d^8y\,\Big(\Big[D^b {\bf V}_y \frac{D_b\bar
D^2}{\partial^2}\phi_y^+ - 4i (\gamma^\mu C)_{bc} \times \nonumber\\
&& \times D^c \bar D^b {\bf V}_y
\frac{\partial_\mu}{\partial^2}\phi_y^+\Big] e^{2V_y}\phi_y -
\mbox{similar terms with $\tilde\phi$}\Big) \Big\rangle\Big|_{p=0}
= 0.\qquad
\end{eqnarray}

\noindent Actually this equality has been already written in Ref.
\cite{Identity} on the language of Feynman diagrams. Nevertheless,
its strict functional formulation is first presented here.

%%%%%%%%%%%%%%%%%%%%%%%%%%%%%%%%%%%%%%%%%%%%%%%%%%%%%%%%%%%%%%%%%%%%

\section{Verification of new identity for Green functions}
\label{Section_Verification}

\subsection{Four-loop approximation in the Abelian case}
\hspace{\parindent}\label{Subsection_Abelian_Verification}

\begin{figure}[h]
    \centering
    \includegraphics[width=0.6in]{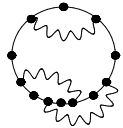}
    \caption{Schematic picture of group of diagrams,
    which is used for verification of new identity}
    \label{explain}
\end{figure}

In order to be sure that the calculations, presented above, and
the proposal about existence of the new identity for Green
functions are true, it is necessary to perform a verification by
explicit calculations. Making this verification is considerably
simplified, if we note that expressions (\ref{First_Diagram}) and
(\ref{Second_Diagram}) allow not only finding sums of all diagrams
with two external lines of the gauge field, but also summing
special classes of such diagrams. Such classes of diagrams are
obtained from a frame, to which external lines are attached by all
possible ways. In particular, in order to verify Eqs.
(\ref{First_Diagram}) and (\ref{Second_Diagram}) in Ref.
\cite{Pimenov} we considered a group of diagrams, which were
obtained from a frame, presented in Fig. \ref{explain}, by
attaching two external line of the gauge field.

This diagrams can be calculated by two different ways:

1. using Eqs. (\ref{First_Diagram}) and (\ref{Second_Diagram}) and
the functions $G$, $f$ and $F$, obtained preliminary.

2. by explicit calculation using supergraphs.

Thus it is possible to perform a four-loop verification of Eqs.
(\ref{First_Diagram}) and (\ref{Second_Diagram}), and also
identity (\ref{Massless_New_Identity1}).

In order to find diagrams which will be essential for obtaining
the unknown functions $G$, $f$ and $F$ in the considered case, it
is convenient to use the following simple speculations: A diagram,
presented in Fig. \ref{explain}, can be considered as a formal
product of the one- and two-loop diagrams with $\phi^*$ and $\phi$
external lines (pairs of their ends are identified):

\begin{figure}[h]
\centering \vspace{0.4cm}
\includegraphics[width=0.6in]{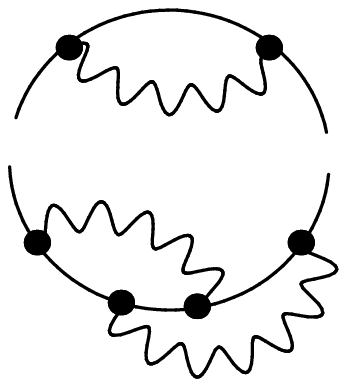}
\end{figure}

\noindent or as the three-loop diagrams with the identified ends:

\begin{figure}[h]
\centering \vspace{0.6cm}
\includegraphics[width=0.6in]{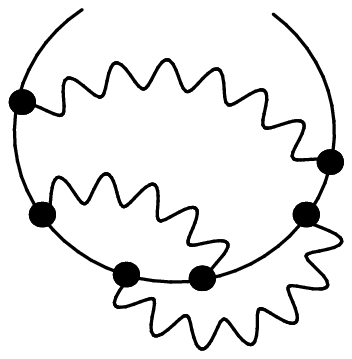} \hspace{2cm}
\includegraphics[width=0.6in]{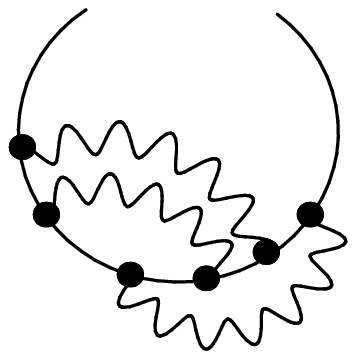}
\end{figure}

\noindent The parts of the diagram, obtained by this way, are the
Feynman diagrams for finding the function $G$. To find the
functions $f$ and $F$, it is necessary to add one more external
line of the superfield $V$ to such diagrams.

\begin{figure}[h]
    \centering
    \includegraphics[scale=0.7]{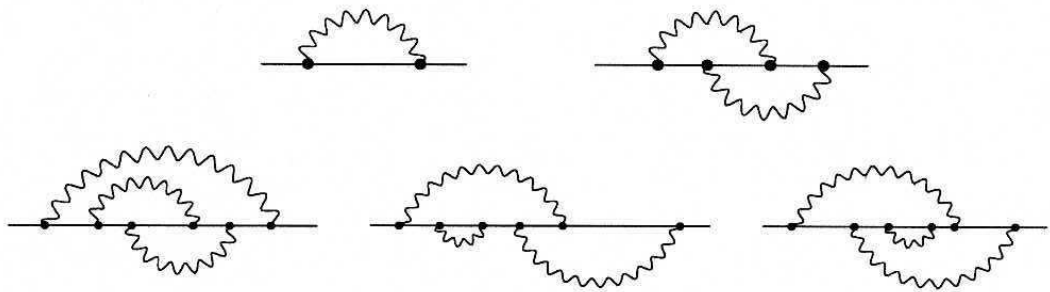}
    \caption{The diagrams, used for calculation of the function $G$}
    \label{1-2-3_loop_anomal}
\end{figure}

Thus, it order to find the function $G$ in the considered case it
is necessary to calculate diagrams presented in Fig.
\ref{1-2-3_loop_anomal}. Then the function $G$ is obtained
according to definition (\ref{Explicit_Green_Functions}). One-,
two- and three-loop parts of the function $G$, defined by diagrams
in Fig. \ref{1-2-3_loop_anomal}, will be denoted by $G_1$, $G_2$
and $G_3$ respectively. Expressions for them, obtained by
calculating these diagrams, are

\begin{eqnarray}\label{G_1}
&& G_1 = -2 e^2 \int \frac{d^4 k}{(2\pi)^4} \,\Big( 1+
\frac{k^{2n}}{\Lambda^{2n}} \Big)^{-1} \, \frac{1}{k^2\,
(q+k)^2}; \\
\label{G_2} && G_2 = - 4 e^4 \int
\frac{d^4k\,d^4l}{(2\pi)^8}\,\Big(1+ \frac{k^{2n}}{\Lambda^{2n}}
\Big)^{-1} \Big(1+ \frac{l^{2n}}{\Lambda^{2n}} \Big)^{-1}
\frac{(2q + k +l)^2}{k^2\, l^2 (k+q)^2\, (q+l)^2\,
(k+q+l)^2};\nonumber\\
\\
\label{G_3} && G_3 = -4\, e^6 \int
\frac{d^4k\,d^4l\,d^4r}{(2\pi)^{12}}\,\Big(1+
\frac{k^{2n}}{\Lambda^{2n}} \Big)^{-1} \Big(1+
\frac{l^{2n}}{\Lambda^{2n}} \Big)^{-1}
\Big(1+ \frac{r^{2n}}{\Lambda^{2n}} \Big)^{-1} \times\nonumber\\
&&\times \frac{1}{k^2 l^2 r^2 (q+k)^2 (q+l)^2 (q+k+l)^2}
\Bigg(\frac{(2q + 2k + r +l)^2 (q+l)^2}{ (q+r+k)^2\,(q+r+k+l)^2}
+\nonumber\\
&& + \frac{(2q + k +l)^2}{ (q+r+k+l)^2} + \frac{2 (2q + k
+l)^2}{(q+r+k)^2} \Bigg).\qquad
\end{eqnarray}

\noindent The complete function $G$ (certainly without diagrams,
which are not essential for this paper) in the considered
approximation is given by

\begin{equation}\label{G}
G(q^2) = 1 + G_1 + G_2 + G_3,
\end{equation}

\noindent where the unity is a tree contribution.

The functions $F(q^2)$ and $f(q^2)$ are defined from three-point
Green functions by Eq. (\ref{Vertex3}). As in the case of the
function $G$, in order to obtain a contribution, which corresponds
to the considered class of the four-loop diagrams, it is
sufficient to calculate only diagrams of a special form. They are
obtained from diagrams, presented in Fig. \ref{1-2-3_loop_anomal},
by all possible insertions of one external $V$-line. In all these
diagrams one external straight line corresponds to the chiral
field $\phi$ and the other corresponds to the non-chiral field
$\phi^*_0$.

We will denote one-, two- and three-loop contributions to the
function $f$  by $f_1$, $f_2$ and $f_3$ respectively. Similar
notation we will use for the function $F$. Because in the tree
approximation these functions are 0 and $f_1=0$, in the considered
order we have

\begin{equation}
F = F_1 + F_2 + F_3;\qquad f = f_2+ f_3.
\end{equation}

\noindent Expressions for $f_1$, $f_2$, $f_3$, $F_1$ and $F_2$,
obtained by calculation of the above pointed diagrams have the
following form:

\begin{eqnarray}\label{f-1}
&& f_1 (q^2) = 0;\vphantom{\Big(}\\
\label{f-2} && f_2(q^2) = \frac{1}{2}\, e^4\int \frac{d^4k \,
d^4l}{(2\pi)^{8}}\,\Big(1+ \frac{k^{2n}}{\Lambda^{2n}} \Big)^{-1}
\Big(1+ \frac{l^{2n}}{\Lambda^{2n}}
\Big)^{-1}\, \times\nonumber\\
&&\quad \quad \times\frac{1}{k^2\, l^2\, (k+q)^2\, (q+l)^2
(k+q+l)^2} \times \nonumber\\
&& \qquad \times \Bigg( -2 + \frac{(2q+k+l)_{\mu}
(k+q)^{\mu}}{(k+q)^2} + \frac{(2q+k+l)_{\mu} (l+q)^{\mu}}{(l+q)^2}
\Bigg);\\
\label{f-3} && f_3(q^2)= 4\, e^6 \int \frac{d^4 k\, d^4 l\, d^4
r}{(2\pi)^{12}}\,\Big(1+ \frac{k^{2n}}{\Lambda^{2n}} \Big)^{-1}
\Big(1+ \frac{l^{2n}}{\Lambda^{2n}}
\Big)^{-1} \Big(1+ \frac{r^{2n}}{\Lambda^{2n}} \Big)^{-1}
\times\nonumber\\
&& \times \frac{1}{k^2\, l^2\, r^2\, (k+q)^2\, (q+l)^2\,
(q+k+l)^2}\,
\Bigg(\frac{1}{(k+q+r)^2} + \frac{1}{2(k+l+q+r)^2} - \nonumber \\
&& - \frac{(2q+k+l)^{\mu} (k+q)^{\mu}}{2(k+q)^2\, (k+q+r)^2} -
\frac{(2q+k+l)^{\mu}
(l+q)^{\mu}}{2(l+q)^2\, (k+q+r)^2}
- \frac{(2q+k+l)^{\mu} (k+q)^{\mu}}{2(k+q)^2\, (k+l+q+r)^2}
- \vphantom{\Bigg(} \nonumber \\
&& - \frac{(2q+k+l)^{\mu} (k+q+r)^{\mu}}{2(k+q+r)^4} \Bigg).
\end{eqnarray}

\begin{eqnarray}\label{F_1}
&& F_1 (q^2) = -\, \frac{e^2}{8} \int \frac{d^4 k}{(2\pi)^4}\,
\Big(1+ \frac{k^{2n}}{\Lambda^{2n}}
\Big)^{-1}\frac{1}{k^2\, (k+q)^4};\\
\label{F_2} && F_2(q^2) = \frac{1}{4}\, e^4\int \frac{d^4k \,
d^4l}{(2\pi)^{8}}\,\Big(1+ \frac{k^{2n}}{\Lambda^{2n}} \Big)^{-1}
\Big(1+ \frac{l^{2n}}{\Lambda^{2n}}
\Big)^{-1}\times\nonumber\\
&&\times \frac{1}{k^2\, l^2\, (k+q)^2\, (q+l)^2 (k+q+l)^2} \Bigg(
4 -2\, \frac{(q+l)^2}{(k+q)^2} -\, \frac{(2q+k+l)^2}{(q+k+l)^2}
\Bigg).
\end{eqnarray}

\noindent An expression for $F_3$ has been calculated, but it is
not presented because it is very large. (It is not required for
verification of identity (\ref{Massless_New_Identity1}).)

Using the obtained expressions it is possible to verify identity
(\ref{Massless_New_Identity1}). With the considered accuracy we
have

\begin{equation}\label{-8fG}
\frac{f}{q^2 G} = \frac{1}{q^2 } \Big(f_3 - f_2\cdot G_1\Big).
\end{equation}

\noindent Substituting here expressions for the functions $G_1$,
$f_2$ and $f_3$ from Eqs. (\ref{G_1}), (\ref{f-2}), (\ref{f-3}) we
obtain, that the integrand can be written as a total derivative
with respect to the momentum $q$

\begin{eqnarray}\label{f-total_deriv}
&& \int \frac{d^4q}{(2\pi)^4}\,\frac{d}{d\ln\Lambda}\,\frac{f}{q^2
G} = e^6 \int \frac{d^4q\, d^4k\, d^4l\, d^4r
}{(2\pi)^{16}}\,\frac{d}{d\ln\Lambda}\Bigg[
\Big(1+\frac{k^{2n}}{\Lambda^{2n}} \Big)^{-1}
\Big(1+\frac{l^{2n}}{\Lambda^{2n}} \Big)^{-1}
\times\qquad\nonumber\\
&& \times \Big(1+ \frac{r^{2n}}{\Lambda^{2n}} \Big)^{-1}\Bigg]
\frac{\partial}{\partial q^{\mu}}\left\{ \frac{(2q+k+l)^{\mu}}{k^2
l^2 r^2 q^2 (k+q)^2 (q+l)^2 (q+r)^2 (k+q+l)^2} \right\}.
\end{eqnarray}

\noindent This equality can be checked by calculating the
derivative with respect to $q^\mu$ using the Leibnitz rule and
comparing the result with Eq. (\ref{-8fG}), in which expressions
for the functions $G_1$, $f_2$ and $f_3$ are obtained by explicit
calculation of diagrams.

Because the integrand in Eq. (\ref{f-total_deriv}) is a total
derivative with respect to $q^{\mu}$ of the expression, which goes
to 0 in the limit $q\to\infty$, expression (\ref{f-total_deriv})
is 0. This means that identity (\ref{Massless_New_Identity1}) is
correct in the considered approximation and for the considered
class of diagrams.

Note that the equalities, present above, allow checking the method
of summing Feynman diagrams by using the Schwinger-Dyson equations
and Ward identities. For this purpose it is possible to calculate
both effective diagrams in Eq. (\ref{SD_Equation}) in the
four-loop approximation explicitly and compare the result with
Eqs. (\ref{First_Diagram}) and (\ref{Second_Diagram}). In
considered approximation

\begin{eqnarray}
&& \frac{1}{q^2}\,\frac{d}{dq^2}\ln(G^2) = \frac{q^{\mu}}{q^4}\,
\frac{\partial}{\partial q^{\mu}}\Big( G_3
-G_1 \cdot G_2 \Big);\\
&& \frac{F}{q^2 G} = \frac{1}{q^2 } \Big(F_3 - F_2\cdot G_1-
F_1\cdot G_2\Big).
\end{eqnarray}

\noindent Explicit calculations were made by the method, proposed
in Ref. \cite{Vestnik}, which simplified finding a part of
diagram, proportional to $V\partial^2 \Pi_{1/2}V$. Nevertheless,
it does not allow to obtain a part, proportional to $V^2$. That is
why the verification of Eqs. (\ref{First_Diagram}) and
(\ref{Second_Diagram}) was made only for terms, proportional to
$V\partial^2\Pi_{1/2} V$. In the both cases this verification
completely confirms them.

\begin{figure}[h]
    \centering
    \includegraphics[width=1.5in]{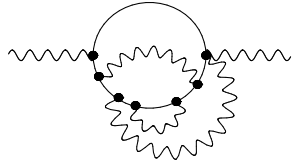}\\
    \caption{One of the diagrams, giving contribution to both
    effective diagrams}\label{Figure_Example}
\end{figure}

As a small technical remark let us note, that making this
verification it is necessary to take into account that a large
number of ordinary Feynman diagrams contributes both to the first
effective diagram in Eq. (\ref{SD_Equation}) and to the second
one. For example, it is easy to see, that a contribution of a
diagram, presented in Fig. \ref{Figure_Example}

\noindent is divided into parts $3/4$ and $1/4$, which correspond
to the first and to the second diagrams in Eq.
(\ref{SD_Equation}).

%%%%%%%%%%%%%%%%%%%%%%%%%%%%%%%%%%%%%%%%%%%%%%%%%%%%%%%%%%%%%%%%%%%%%%%%%

\subsection{Three-loop approximation in the non-Abelian case}
\hspace{\parindent}\label{Subsection_Nonabelian_Verification}

\begin{figure}[h]
    \centering
    \includegraphics[scale=0.7]{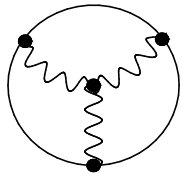}
    \caption{Diagram, generating considered contribution to the
    new identity}
    \label{ostov}
\end{figure}

The Feynman rules are different in a non-Abelian theory mostly due
to vertexes with the selfaction of the gauge field. That is why it
is desirable to check the new identity also in this case
\cite{3ident}. For diagrams that do not contain such vertexes the
calculations are similar to the Abelian case. But for diagrams
with the triple vertex of the gauge field an additional
verification is very desirable. As we already mentioned, for this
purpose it is not necessary to calculate all Feynman diagrams in a
given order of the perturbation theory. It is sufficient to
consider, for example, a typical three-loop diagram, presented in
Fig. \ref{ostov}.

\vspace{0.5cm}

\vspace{0.5cm}
\begin{figure}[h]
    \centering
    \includegraphics[scale=0.7]{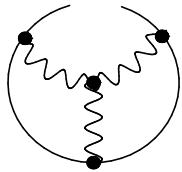}
    \caption{Way of cutting the diagram}
    \label{ostov-cut}
\end{figure}

\begin{figure}[h]
    \centering
    \includegraphics[scale=0.65]{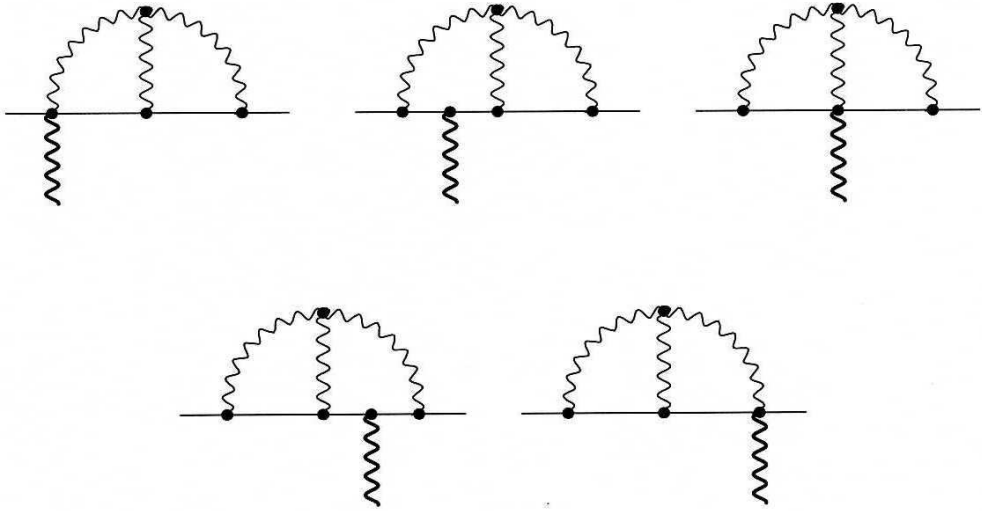}
    \caption{Diagrams, defining the function $f$, corresponding to
    the considered diagrams}
    \label{vertex-diagr}
\end{figure}

\noindent From the topological point of view there is the only way
to cut a loop of the matter superfield, presented in Fig.
\ref{ostov-cut}. Hence, it is necessary to calculate a set of
diagrams, presented in Fig. \ref{vertex-diagr}. As earlier, in all
these diagrams the chiral field $\phi$ is at the first external
line and the non-chiral field $\phi^*_0$ is at the second line.
Therefore, all presented diagrams are not topologically
equivalent.

\vspace{0.5cm}

Calculating these diagrams we can find the function $f$. The
function $G$ in the lowest approximation should be set to 1.
Really, in the tree approximation $G=1$. Hence, in the given order
for the considered class of diagrams we have:

\begin{equation}
G(q^2) = 1 + O(\alpha^2);\qquad f(q^2) = f_2(q^2) + O(\alpha^3).
\end{equation}

\noindent where $f_2$ is proportional to $\alpha^2$. Therefore,

\begin{eqnarray}\label{Expansion0}
\int d^4q\,\frac{d}{d\ln\Lambda}\frac{f(q^2)}{q^2 G(q^2)} = \int
d^4 q\,\frac{d}{d\ln\Lambda} \frac{f_2(q)}{q^2} + O(\alpha^3).
\end{eqnarray}

\noindent So, we see that the considered contribution is actually
determined by the two-loop value of the single function $f_2$.

In order to find the function $f_2$ in the two-loop approximation,
it is necessary to make an explicit calculation of Feynman
diagrams, presented in Fig. \ref{vertex-diagr}, using the standard
supergraph technique. The result is (in the Euclidean space, after
the Weak rotation)

\begin{eqnarray}\label{f-expression}
&& f_2(q) = - 2\pi^2 \alpha^2 C_2 \Big( C_2(R) - \frac12\, C_2
\Big)\, \int \frac{d^4k\, d^4l}{(2\pi)^8} \Bigg(
\frac{l^{\mu}}{(k+q+l)^2} + \frac{(k+q)^{\mu}}{(k+q)^2}
\Bigg)\times\\
&& \times \frac{(k+q+l)_{\mu}}{(k+q)^2 \, (k+q+l)^2 k^2 \Big(1 +
k^{2n}/\Lambda^{2n}\Big)\, l^2 \Big(1 + l^{2n}/\Lambda^{2n}\Big)\,
(k+l)^2 \Big(1 + (k+l)^{2n}/\Lambda^{2n}\Big)},\nonumber
\end{eqnarray}

\noindent where $C_2(R)$ and $C_2$ are defined by

\begin{equation}\label{C1-inv}
T^a\, T^a = C_2(R),
\end{equation}
\begin{equation}\label{C2-inv}
f^{amn} \, f^{bmn} = C_2 \, \delta^{ab}.
\end{equation}

\noindent Substituting this expression into Eq.
(\ref{Expansion0}), we obtain that in the considered approximation
for the considered diagrams

\begin{eqnarray}
&& \int
\frac{d^4q}{(2\pi)^4}\,\frac{d}{d\ln\Lambda}\frac{f(q^2)}{q^2\,G(q^2)}
=\nonumber\\
&& = \alpha^2 \pi^2 \, C_2 \Big( C_2(R) - \frac12\, C_2 \Big)\,
\int \frac{d^4q\, d^4k\, d^4l}{(2\pi)^{12}}
\,\frac{\partial}{\partial q^{\mu}}
\Bigg\{\Lambda\frac{d}{d\Lambda} \, \Bigg[
\frac{(k+q+l)^{\mu}}{q^2 \, (k+q)^2 \,
(k+q+l)^2} \times\qquad \nonumber\\
&& \times \frac{1}{k^2 \Big(1 + k^{2n}/\Lambda^{2n}\Big) \, l^2
\Big(1 + l^{2n}/\Lambda^{2n}\Big)\, (k+l)^2 \Big(1 +
(k+l)^{2n}/\Lambda^{2n}\Big)}\Bigg] \Bigg\} = 0.
\end{eqnarray}

\noindent This means that the new identity for Green functions is
also valid in the non-Abelian theory.

%%%%%%%%%%%%%%%%%%%%%%%%%%%%%%%%%%%%%%%%%%%%%%%%%%%%%%%%%%%%%%%%%%%

\section{Conclusion}
\label{Section_Conclusion}
\hspace{\parindent}

Our investigation of applying the higher derivative regularization
to calculation of quantum corrections in supersymmetric theories
allows revealing some interesting features, which were not noted
earlier. We should first mention the completely unexpected result:
all complicated integrals, defining the Gell-Mann--Low function,
are integrals of total derivatives and can be easily calculated.
It is important to note that with the higher derivative
regularization the Gell-Mann--Low function is calculated in the
simplest way. According to our calculations this is a function
that coincides with the exact NSVZ $\beta$-function. So, it is not
necessary to tune a renormalization scheme, as it was done
earlier, calculating the $\beta$-function defined by divergence in
the $\overline{MS}$-scheme with the dimensional reduction. Note
that with the higher covariant derivative regularization there are
divergences only in the one-loop approximation. In a certain
degree this confirms speculations, presented in Ref. \cite{SV}, in
which the authors suggested that the Wilsonian action was
exhausted at the one-loop. The calculations show that with the
higher derivative regularization the renormalized action is
exhausted at the one-loop.

The features, which appear calculating quantum corrections in
supersymmetric theories, can be partially explained by
substituting solutions of the Slavnov--Taylor identities into the
Schwinger--Dyson equations. However, it is necessary to suppose
existing a new identity for the Green functions, which does not
follow from known symmetries of the theory. The exact
$\beta$-function can be also considered as a similar identity.
Quite possible that all such identities are consequences of some
nontrivial symmetry. Existence of such symmetries was already
proposed earlier \cite{Ermushev} in $N=1$ finite supersymmetric
theories. That is why the obtained results could be also useful
for their investigation. In particular, it is possible that the
anomalous dimension in $N=1$ finite theories is also an integral
of a total derivative. Now this statement is being checked by the
explicit calculation.

Finally, it is necessary to enumerate some interesting problems,
which have not yet been solved. One of them is a proof of the new
identity starting from some symmetry. May be, there is a relation
between the considered problem and the AdS/CFT-correspondence. In
favour of this we note that it is convenient to formulate the new
identity in terms of vacuum expectation values of gauge invariant
operators. Possibly it would be possible to derive the NSVZ
$\beta$-function for the pure Yang--Mills theory exactly to all
orders of the perturbation theory. In this case the
Schwinger--Dyson equations are rather involved and their
investigation is much more complicated.

To conclude, we can say that dynamics of supersymmetric theories
is unexpectedly rich by the interesting features, which sometimes
are hard to be explained. The higher covariant derivative
regularization is an excellent tool for revealing these features.

\bigskip
\bigskip

\noindent {\Large\bf Acknowledgments.}

\bigskip

\noindent This paper was partially supported by the Russian
Foundation for Basic Research (Grant No. 05-01-00541).

%%%%%%%%%%%%%%%%%%%%%%%%%%%%%%%%%%%%%%%%%%%%%%%%%%%%%%%%%%%%%%%%%%%%%%%%%

\end{document}